\documentclass[manuscript]{acmart}

\usepackage{enumitem}
\usepackage{booktabs}
\usepackage{tabularx}
\usepackage{xcolor,colortbl}
\usepackage{tabularx}
\definecolor{lightgray}{gray}{0.93}
\definecolor{slightgray}{gray}{0.98}
\definecolor{darkgray}{gray}{0.77}
\usepackage{makecell}
\usepackage{multirow}
\usepackage{threeparttable}
\usepackage{hyperref}
\usepackage{graphicx}
\RequirePackage[normalem]{ulem} 
\RequirePackage{color}\definecolor{RED}{rgb}{1,0,0}\definecolor{BLUE}{rgb}{0,0,1} 

\newcolumntype{Y}{>{\centering\arraybackslash}X}
\AtBeginDocument{%
  \providecommand\BibTeX{{%
    \normalfont B\kern-0.5em{\scshape i\kern-0.25em b}\kern-0.8em\TeX}}}

\begin{document}

\title{CoRemix: Supporting Informal Learning in Scratch Community With Visual Graph and Generative AI}

\author{Yunnong Chen}
\affiliation{%
  \institution{Zhejiang University}
  \city{Hangzhou}
  \country{China}}
\email{chen_yn@zju.edu.cn}

\author{Yishu Shen}
\affiliation{%
  \institution{Shanghai Jiao Tong University}
  \city{Shanghai}
  \country{China}}

\author{Ruiyi Liu}
\affiliation{%
  \institution{Tongji University}
  \city{Shanghai}
  \country{China}}

\author{xinyu Yu}
\affiliation{%
  \institution{Zhejiang University}
  \city{Hangzhou}
  \country{China}}
\email{3210104724@zju.edu.cn}

\author{Lingyun Sun}
\affiliation{%
  \institution{Zhejiang University}
  \city{Hangzhou}
  \country{China}}
\email{sunly@zju.edu.cn}

\author{Liuqing Chen}
\authornote{Corresponding author.}

\affiliation{%
  \institution{Zhejiang University}
  \city{Hangzhou}
  \country{China}}
\email{chenlq@zju.edu.cn}

\renewcommand{\shortauthors}{Chen et al.}

\begin{abstract}
Online programming communities provide a space for novices to engage with computing concepts, allowing them to learn and develop computing skills using user-generated projects. However, the lack of structured guidance in the informal learning environment often makes it difficult for novices to experience progressively challenging learning opportunities. Learners frequently struggle with understanding key project events and relations, grasping computing concepts, and remixing practices. This study introduces CoRemix, a generative AI-powered learning system that provides a visual graph to present key events and relations for project understanding. We propose a visual-textual scaffolding to help learners construct the visual graph and support remixing practice. Our user study demonstrates that CoRemix, compared to the baseline, effectively helps learners break down complex projects, enhances computing concept learning, and improves their experience with community resources for learning and remixing.
\end{abstract}

\ccsdesc[500]{Social and professional topics~Computational thinking}
\ccsdesc[500]{Social and professional topics~K-12 education}
\ccsdesc[500]{Human-centered computing~Interactive systems and tools}
\keywords{Online Community, Informal Learning, Programming Learning, Generative AI}



\maketitle

\section{Introduction}

\begin{figure*}[t]
    \centering
    \includegraphics[width=1\textwidth]{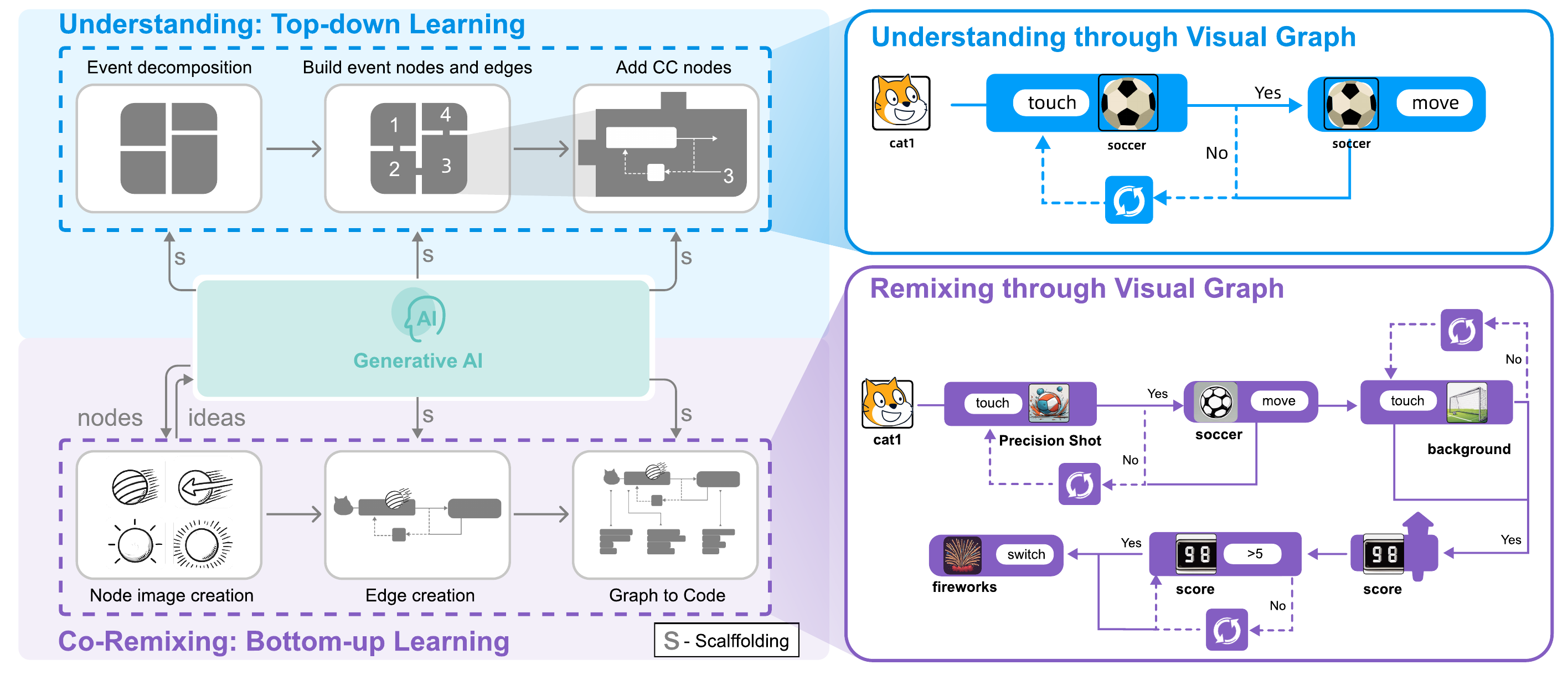}
    \caption{\textbf{Overview of \textit{CoRemix}. In the understanding phase, learners decompose project events, create event nodes and edges, and incorporate computing concept (CC) nodes to understand key computing concepts, guided by generative AI scaffolding. In the co-remixing phase, learners engage in remixing, adding new nodes and relationships to enhance their projects.}}
    \vspace{-0.25in}
    \label{fig:overview}
\end{figure*}

Recent research increasingly views interest-driven online communities as promising environments for supporting learners \cite{bruckman1998community, cheng2020building, cheng2020critique}. Informal learning in such communities refers to self-directed, unstructured learning that takes place within collaborative settings, such as online communities \cite{monroy2007scratchr} or social platforms \cite{barrot2022social}. In these environments, community members share the artifacts they create and engage in discussions with others \cite{dasgupta2017scratch, dasgupta2016remixing}, often without formal guidance or curricula. Informal learning communities, such as Scratch online communities, encourage creativity and peer collaboration \cite{burke2016computational, foong2017online, ford2018we}, allowing learners to develop skills through hands-on practice and interaction with others.

Although online informal learning environments like the Scratch community offer rich opportunities for learners to engage with programming projects, the lack of sufficient guidance can lead to uneven knowledge acquisition and an overrepresentation of certain ways of applying knowledge, potentially hindering creative freedom \cite{cheng2022interest}. Quantitative analyses of the Scratch online community have shown that users who remix or build upon others' code often lack innovation, raising concerns about their development of transformative abilities \cite{matias2016skill, yang2015uncovering, cheng2022many}. Most Scratch users exhibit only a limited range of programming skills, suggesting that without the support of experienced teachers or structured classroom environments, beginners struggle to achieve significant advancements through community-based informal learning \cite{hill2013remixing, dietz2023visual, chen2024chatscratch}. This evidence highlights the need for better informal learning support tools to facilitate interest-driven learning in online programming communities.

In this study, we target novice young programmers who primarily consume user-generated programming projects in online communities as learning materials but do not engage in project creation. Our goal is to leverage the rich programming resources available in these communities to support beginners in learning computing concepts (CC) and engaging in computing practices (CP). On one hand, we recognize that interesting and simple projects can serve as effective starting points for novices to construct their own projects and express creative ideas, aligning with the Zone of Proximal Development (ZPD) \cite{chaiklin2003zone, chen2024chatscratch, dietz2021storycoder}. On the other hand, the complexity and high integration of many community projects make it challenging for novices to fully grasp the underlying computing concepts.

To understand the specific challenges learners face in community-based informal learning, we conducted a formative study with six young novice learners and four Scratch education experts. Through observations of the learners' processes and interviews with the experts, we identified several key challenges: difficulties in understanding the execution flow and logical relations in block-based programming projects; challenges in learning computing concepts from complex project codes; and difficulties in applying creative programming ideas to projects of personal interest. These findings underscore the necessity of providing targeted support to help novices navigate and learn from the wealth of resources in online programming communities.

To address these challenges, we introduced CoRemix (Figure \ref{fig:overview}), an innovative interactive system that seamlessly integrates visual graph representations with generative AI. The visual graph is composed of event nodes and computing concept nodes to represent programming projects. CoRemix employs a unique combination of top-down and bottom-up learning approaches: it initially guides learners through event decomposition to construct events and relationships, and then supports them in creating new event nodes, establishing connections, and completing the visual graph by understanding the project. We build a conversational agent powered by generative AI, designed to guide learners in project understanding and remixing practices. This agent provides personalized scaffolding, helping novices navigate complex block-based programming projects. We adopt a retrieval-augmented large language model (RA-LLM) to improve the relevance, richness, and educational quality of the agent's responses compared to standard LLM.

To evaluate the effectiveness of CoRemix, we conducted a between-subjects study with 16 children aged 9-12, comparing their use of CoRemix to the Scratch community website as a baseline. The results showed a significant improvement in participants' ability to understand community projects, reflected in the increased identification of key events and relationships, as well as expert ratings. Computational Thinking (CT) scores and measures of remixing activity quality further demonstrated CoRemix's effectiveness in understanding computing concepts and promoting computing practice. Survey results and participant feedback indicated that CoRemix was more effective than the baseline. In the discussion, we explored the various advantages and emerging challenges of using generative AI to support informal learning in communities, analyzed the integration of resource use in visual programming environments, and identified current limitations and future research directions.

This study makes the following contributions:

\begin{itemize}
    \item CoRemix is an innovative interactive system that leverages visual graphs to support project understanding and the learning of computing concepts within the Scratch online community. By using graphical metaphors, it enables the abstraction and decomposition of projects into manageable parts, making complex programming concepts more accessible to learners.
    \item A LLM-based conversational agent is integrated into the system, using a visual-textual prompting pipeline to stimulate learners' critical thinking and provide inspiration for remixing projects. The inclusion of retrieval-augmented LLMs (RA-LLMs) enhances the reliability and educational value of the agent's responses, ensuring more effective guidance.
    \item A comprehensive evaluation, including a technical assessment and a user study with 16 participants, demonstrates that CoRemix significantly improves learners' understanding of projects and engagement in informal learning activities, highlighting its potential to enhance educational experiences in online programming communities.
\end{itemize}

\section{Related Work}

\subsection{Informal Learning in Scratch Community}
Online communities are increasingly becoming learning environments driven by participation, interest, and community support \cite{peng2024designquizzer, guo2023makes,peng2020exploring}. A variety of theoretical frameworks have been employed to design and analyze these communities, many of which draw from foundational theories on the social origins of learning, such as those by Vygotsky \cite{chaiklin2003zone} and Lave \& Wenger \cite{lave1991situated}. In interest-driven online communities, learning primarily occurs through two pathways: the sharing of creative works, such as fanfiction \cite{campbell2016thousands}, design models \cite{cheng2020critique}, or interactive computer programs \cite{dasgupta2016remixing}, and social interaction around these works, including comments, remixes, and critiques. To support the first pathway, many communities are structured to make learning outcomes visible as public artifacts, serving as examples and sources of inspiration for others \cite{dasgupta2016remixing, gan2018gender}, while also acting as scaffolds for replication, practice, and innovation \cite{tausczik2017share}. The Scratch community, for example, incorporates a remixing feature—allowing users to take an existing project and modify it to create their own project—enabling novices to quickly leverage community resources for computing practice. To foster the second pathway, communities offer direct peer-to-peer support, such as comments \cite{wang2021cass} and Q\&A discussions \cite{tausczik2014collaborative}, which help members deepen their understanding of specific topics or techniques \cite{shorey2021hanging}. 

In the Scratch community, learners often engage in social interactions around shared projects and seek feedback from peers and experienced users. However, despite its potential, Scratch community learners in informal settings still face several challenges. Recent studies on online programming communities have highlighted disparities in participation and learning outcomes based on gender and race \cite{fields2014programming, richard2016blind, gan2018gender}. Additionally, understanding key computing concepts or engaging in meaningful computational practices is not always effectively supported through social interactions alone. While online communities provide opportunities for users to draw inspiration from shared examples, research on remixing has suggested that it can sometimes hinder originality \cite{cheng2022interest, hill2013remixing}. This indicates that informal learning activities, such as remixing, without structured guidance, may not always promote computational thinking or creative expression. In this study, we focus on enhancing learning outcomes and experiences by providing structured guidance, from understanding computing concepts to creative programming implementation, through collaborative remixing activities with learners.

\subsection{Graph-based Tools for Computing Education}
Graph-based tools have emerged as a powerful approach to support learning in computing education, offering visual representations that help learners structure and organize their understanding of computing concepts, such as mind maps \cite{lin2022development} and concept maps \cite{chen2020effects, resch2024overcoming}. By using graphs to map out code structures, dependencies, or logical flows, learners can better grasp abstract concepts and relationships \cite{chiou2017analyzing}. This method aligns with cognitive theories of learning, which suggest that visualizations can reduce cognitive load and aid in the comprehension of complex systems \cite{sweller1988cognitive}. Many visual programming tools can be used to support the development of K-12 learners’ CT skills. For example, Scratch uses a block-based programming language, allowing learners to create various types of projects such as stories, games, and animations. A Scratch project can be represented as a combination of events (such as character actions or user inputs) and the relationships between events (such as sequential execution, conditional statements, or loops). However, research has shown that beginners still find it challenging to learn block-based visual programming tools, especially when it comes to understanding complex projects. These environments often lack features like code navigation and explanations, which are crucial for novice learners. In Coremix, we drew inspiration from CodeOrama \cite{ladias2021codeorama}, a tool that visualizes complex Scratch code in a two-dimensional table, helping students and teachers understand program structure, flow, and interactions. While CodeOrama offers valuable insights into code flow visualization, its complex creation process may hinder novice users. In Coremix, we aim to provide a more streamlined form of process visualization to enhance project understanding. 

Moreover, previous studies have shown that graph-based visualization tools can facilitate learners’ creativity \cite{malycha2017enhancing, yan2023xcreation, sun2022students}. In computing education, leveraging creativity support tools to inspire learners' creativity has been shown to increase immersion and motivation by allowing learners to explore and express their ideas more freely. Studies suggest that tools which provide scaffolding for creative tasks can lead to higher levels of engagement and deeper learning outcomes \cite{chen2024chatscratch, zhang2024mathemyths}. Echoing the idea of technology-enhanced creativity championed by Dietz et al. \cite{dietz2021storycoder,dietz2023visual}, in CoRemix we explored a graph-based approach to support learners' creative programming, using generative AI to reduce the burden of multitasking and enhance learner engagement.

\subsection{Support Informal Learning through Learner-AI Collaboration}
In recent years, many researchers have focused on integrating LLMs into educational technology to support informal learning through collaboration between learners and LLMs \cite{kasneci2023chatgpt}. These applications include creating mathematical stories with children using LLMs \cite{zhang2024mathemyths}, employing LLMs as teachable agents to support learners in building metacognitive skills \cite{jin2024teach}, serving as code assistants in classrooms \cite{kazemitabaar2024codeaid}, and creating personalized, creative programming projects \cite{chen2024chatscratch}. For instance, LLMs have been used to generate events and characters in stories, highlighting their powerful capabilities in supporting creativity in programming \cite{chen2024chatscratch}. In different research trajectories, some scholars have utilized LLMs to create intelligent learning partners that collaborate with humans \cite{ji2023systematic}, provide feedback \cite{jia2021all}, and encourage students \cite{gabajiwala2022quiz, tai2023impact}. For example, Jin et al. \cite{jin2024teach} used LLMs as teachable agents to stimulate learners' thinking by generating Socratic questions, shifting away from the traditional question-answer mode of teachable agents. ChaCha \cite{seo2024chacha} introduced a state machine-based prompting pipeline, demonstrating excellent contextual memory capabilities in emotional learning collaboration with children. In the realm of informal learning within communities, DesignQuizzer \cite{peng2024designquizzer} built a quiz pool based on community resources, using language models to generate design quizzes and provide structured feedback. However, existing studies have not explored how to use LLMs to support learners in achieving better learning outcomes and experiences in informal community learning, nor have they fully leveraged community resources to construct educationally meaningful LLMs.
\section{Formative Investigation}

In this section, we conducted a formative study with two main goals: (1) to understand the challenges faced by novice learners during informal learning in the Scratch community, and (2) to explore the effectiveness of LLMs in guiding novice learners. The study consisted of a learning task and a semi-structured interview. 

\subsection{Participants and Procedure}

ten participants were recruited from a specialized institution focused on programming education for young learners, including six novice learners and four experienced educators. The novice learners (L1–L6) were elementary school students aged 9 to 12 years (M = 10.37, SD = 1.05), each having less than one year of experience within the Scratch community. The educators (T1–T4), aged 28 to 30 years (M = 29.25, SD = 0.83), had rich experience in programming education, specifically in Scratch.

In the study, we first conducted a 60-minute learning task. During the initial 10 minutes, novices were required to explore the Scratch community \footnote{\url{https://scratch.mit.edu/}} to select one project. In the following 50 minutes, they were asked to learn independently and remix the selected project, then submit their remixing project. To investigate the effectiveness of LLMs in novices' informal learning, participants were allowed to seek guidance from ChatGPT using an account provided by researchers. Participants were also required to share their screens to facilitate real-time observation, and screen recordings were captured with their consent. Upon completing the task, participants were required to verbalize their understanding of the project. We then conducted 40-minute semi-structured interviews with both the learners and the educators. The interviews with the learners focused on the challenges encountered during the process, while the interviews with the educators centered on potential strategies to address these challenges and their perspectives on the effectiveness of LLMs in guiding novices. With their consent, all interviews were audio-recorded. Detailed interview questions are provided in Appendix A of the supplementary materials.

\subsection{Findings}
\textbf{Challenge 1: Understanding of Key Project Events and Relations.} 
Through interviews and video recordings, we found that novices often exhibit an unstructured learning process. It means that beginners tend to approach learning without a clear plan. Traditional programming typically begins with a clear task and a linear thought process, such as writing a simple loop to print even numbers in Python. In contrast, understanding Scratch projects requires going through a phase where learners identify events and relations occurring in the stage area (visual area). We observed that learners often struggle to break down the project into a combination of events and relations or are only able to identify a few obvious ones \cite{10.1145/1592761.1592779}. However, all six educators emphasized the importance of this phase, noting that children in informal learning settings need support to complete this critical step for better learning outcomes. As L3 mentioned, ``When I started learning the adventure project \footnote{\url{https://scratch.mit.edu/projects/27041093}}, I wanted to understand how to implement the flying action and the relation between the cat and the item, but I didn’t know where to start.'' We observed that although L3 understood that the character could walk and fly, they couldn’t grasp the key conditions and relationships needed to implement those functions. This can be attributed to most children's cognitive development stage, where abstract thinking is still lacking. While two-thirds of the participants expressed enjoyment in the community projects, they struggled with the learning process, leading to more passive learning. To address this issue, two educators suggested using tools like flowcharts and mind maps to provide project breakdowns and reframe the projects in a structured visual format.

\textbf{Challenge 2: Learning computing Concepts in Projects.} The second major challenge learners face in community-based informal learning is understanding the computing concepts embedded in projects. The programming process in Scratch is akin to building with LEGO blocks, where each code segment functions as a distinct building block that can be combined to create more complex features. While learners are accustomed to using computing concepts like loops and variables in formal programming settings, they often encounter complex code blocks in community projects that involve multiple intertwined concepts, such as nested blocks with loops, conditions, and Boolean values. T3 mentioned that even learners with over a year of Scratch experience struggle to comprehend the intricate functionality found in these community projects. In informal learning, learners frequently turn to other community members for assistance, yet this help is often delayed or insufficient. With the advent of ChatGPT, three learners reported using conversational interfaces like ChatGPT to understand the computing concepts behind these complex functionalities. However, we found that mastering the use of LLMs, such as crafting prompts to effectively articulate their needs, remains a significant challenge for beginners. Due to their cognitive development stage, novice learners often find it difficult to provide detailed project context, limiting ChatGPT’s ability to deliver precise answers. Additionally, during interviews, T2 described a step-by-step scaffolding method used in her teaching, where she first identifies the key code blocks contributing to a function and, after allowing students time to reflect, explains the specific role of each block. Both T3 and T4 noted that this type of support was effective, helping them overcome initial hesitation and motivating them to think more critically. These findings inspired improvements in CoRemix, prompting us to design the system to offer multi-step scaffolding strategies that guide beginners toward active reflection while reducing the burden of providing detailed descriptions.

\textbf{Challenge 3: Achieving Creative Remixing Practices is Difficult.} 
Scratch provides multiple ways to create assets, such as importing external assets or using the built-in asset library. However, in practice, children often use only the resources from the original project during remixing. We found that learners are motivated to remix projects they are interested in, and community projects offer them ample inspiration. However, their lack of asset creation skills and sufficient programming knowledge makes it difficult for them to translate their ideas into modular code blocks (L2, L3). Participants noted that they didn’t know how to integrate their ideas into the project, nor how to implement them through programming. In practice, educators often use text-to-image models like Midjourney \footnote{\url{https://www.midjourney.com/}} to provide children with the creative resources they need. However, T4 pointed out that in informal learning environments, such as home-based learning, children may struggle to access similar support and face risks of exposure to harmful content. Based on these findings, we designed a creative support feature for CoRemix that allows learners to create image nodes and relationships through narration during visual graph creation, while ensuring safety through technical safeguards \cite{Kapoor2023NegativePrompts, Stone2023NegativeAIPrompts}. Considering learners' limited programming skills, we continue to provide scaffolding in the graph-to-code process to guide them effectively in implementing their creative ideas through programming.

\subsection{Design Goal}

Based on our formative study, we outline three design goals for the development of CoRemix. Our system aims to assist young learners in effectively understanding and remixing artifacts shared by others within computing learning communities.

\begin{itemize}
    \item \textbf{DG1. Abstract community projects into key events and relations through straightforward visualization.} We found that learners often struggle to understand community projects and are unsure where to begin, which leads them to interact with the projects passively. CoRemix should provide novices a clear starting point, enabling learners to understand key events and relationships.
    \item \textbf{DG2. Support learners in learning computing concepts through step-by-step scaffolding.} CoRemix should provide step-by-step scaffolding to encourage learners to actively engage with the logic and computing concepts. This deeper form of engagement is more effective for learning than passive activities such as merely reading or basic manipulations (e.g., adjusting loop counts or switching images).
    \item \textbf{DG3. Provide interactive guidance during remixing practices, allowing learners to practice within the community.} We found that learners struggle to achieve creative remixing practice due to a lack of sufficient programming and material support. CoRemix should help novices effectively explore personal meaningful ideas and engage these in the learning process through programming.
\end{itemize}

\section{CoRemix}
Based on three design objectives, we developed CoRemix, which follows a two-phase process: project understanding and collaborative remixing (Figure \ref{fig:ui}). The goal of the project understanding phase is for learners to construct a visual graph by breaking down key event nodes and establishing logical relationships between them. The visual graph also integrates computing concept nodes to represent the computational thinking involved in the project. In the remixing phase, the system supports learners' creativity during remixing activities. To facilitate this, we developed a conversational agent that employs RA-LLMs and text-to-image models to assist learners in building visual graphs and exploring remixing ideas. This section details CoRemix’s design principles and user flow.

\subsection{Visual Graph}

\begin{figure*}[htp]
    \centering
    \includegraphics[width=0.9\textwidth]{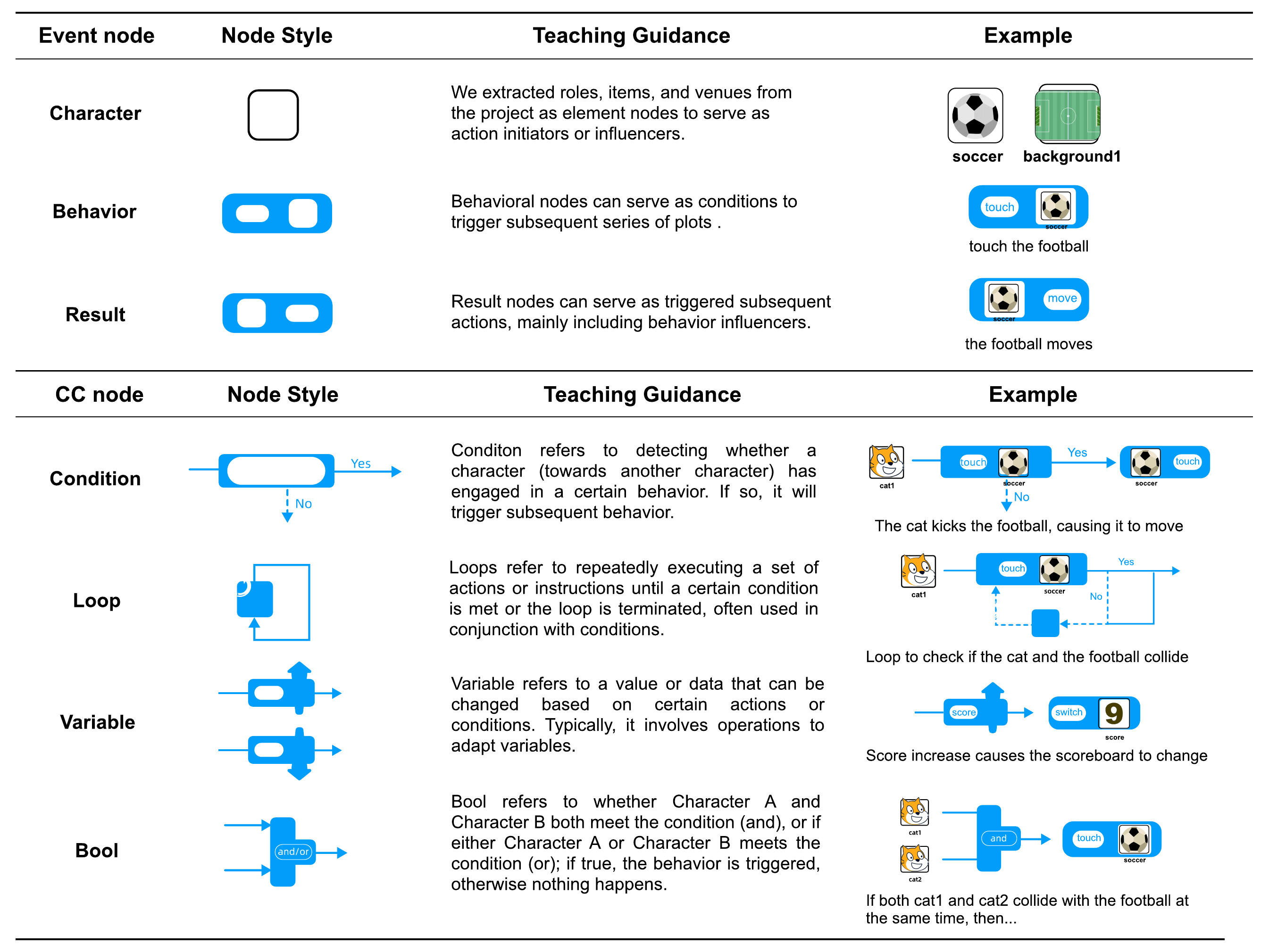}
    \caption{\textbf{Event and CC Nodes: In \textit{CoRemix}, the child can use two types of nodes to build \textit{visual graph}: Event nodes (Character, Behavior, and Result) and CC nodes (Condition, Boolean, Loop, and Variable).}}
    \vspace{-0.11in}
    \label{fig:node}
\end{figure*}

Our goal is to provide an abstract representation of the project, allowing learners to build their visual graph based on this abstraction.  To achieve this, we defined two main node types: event nodes and computing concept nodes. Event nodes represent key programming actions such as character behaviors and outcomes while computing concept nodes encapsulate fundamental programming logic like conditions, loops, variables, and booleans. We scraped 5,000 projects from four categories in the Scratch community (games, animations, stories, and math) and used static analysis to determine that these computing concepts form the basis for representing most projects, accounting for 21\%, 30\%, 24\%, and 7\% of the blocks, respectively. According to Figure \ref{fig:node}, the nodes in the visual graph can be one of the following:

\begin{itemize}[label=$\ast$]
    \item \textit{Event node}: Represent the actions and outcomes of characters within a project. These nodes abstractly capture the functions performed by code blocks and their visual effects. Learners can drag and drop event nodes onto the visual graph and connect them using relationships established by the computing concept nodes.
    \item \textit{CC nodes}: Represent core computing concepts necessary for implementing a character’s actions in the event, such as conditions, loops, and variables. These nodes allow learners to represent the computing concepts in the visual graph. 
\end{itemize}

\subsection{Abstracting Project to Visual Graph}

\begin{figure*}[htp]
    \centering
    \includegraphics[width=0.95\linewidth]{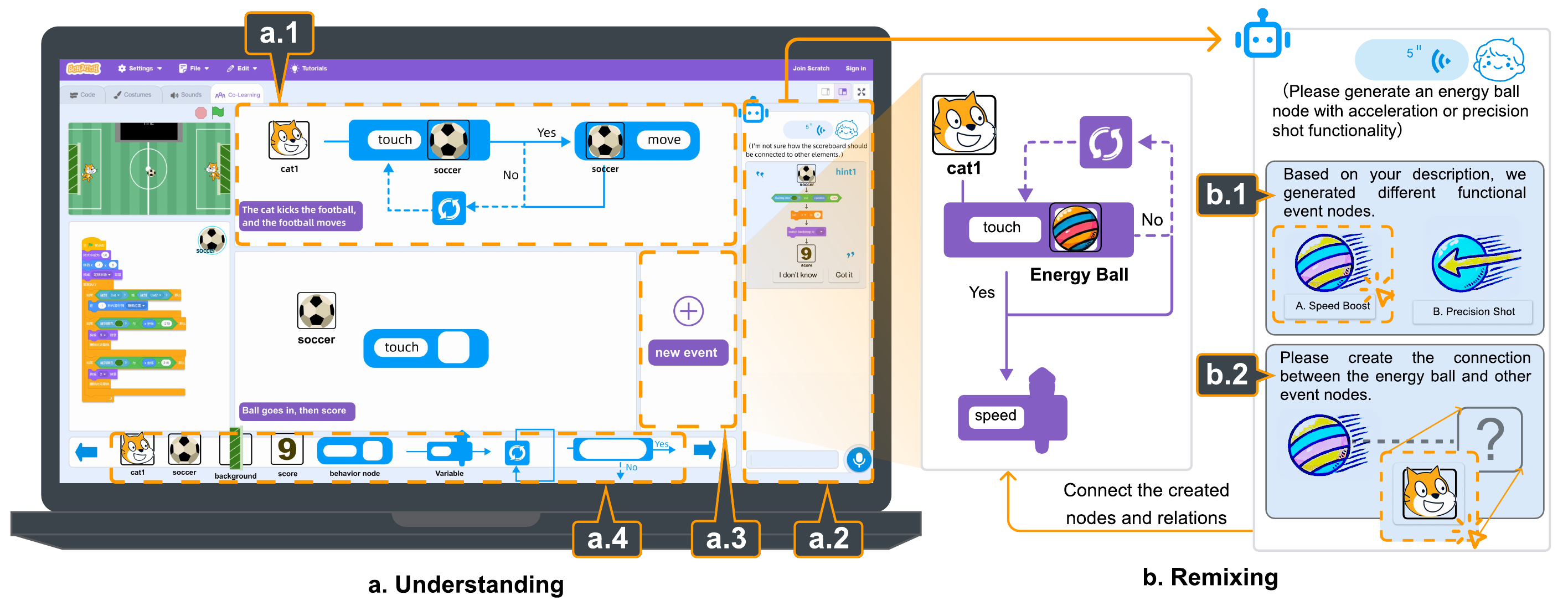}
    \caption{\textbf{When using \textit{CoRemix}, learners first create event and computing concept nodes from the node area (a.4) and build edges on the visual graph (a.1). Learners can also add corresponding event descriptions on the canvas. If they encounter difficulties while constructing the graph, they can get scaffolding support from the conversational agent equipped with generative AI (a.2). Then, learners can click ``new event'' (a.3) to create a new canvas and start remixing the project by generating new event nodes with images (b.1) and building relationships with existing nodes (b.2).}}
    \vspace{-0.21in}
    \label{fig:ui}
\end{figure*}

Based on interviews, we aim to help learners abstract the community project into key programming events and relationships through direct visualization, integrating computing concepts such as sequences, loops, events, and variables \textbf{(DG1)}. In the understanding phase, learners first construct the visual graph by using the character nodes provided by the system (a.4)). Learners are then guided to connect the event and character nodes to represent the logical sequence of the project. We provide a chat window for learners to interact with CoRemix (Figure \ref{fig:ui}.(a.3)). If the user is unsure how to build the visual graph, CoRemix provides a hint to assist in event decomposition (e.g., ``Let’s think together. Do you think there might be a connection between scoring a goal and the score?''). Once learners feel they have completed the decomposition, they can click the ``new event'' button to create a canvas for a visual graph (Figure \ref{fig:ui}.(a.1)).

\subsection{Learning computing Concepts through Visual-Textual Scaffolding}
\begin{figure*}[htp]
    \centering
    \includegraphics[width=0.95\textwidth]{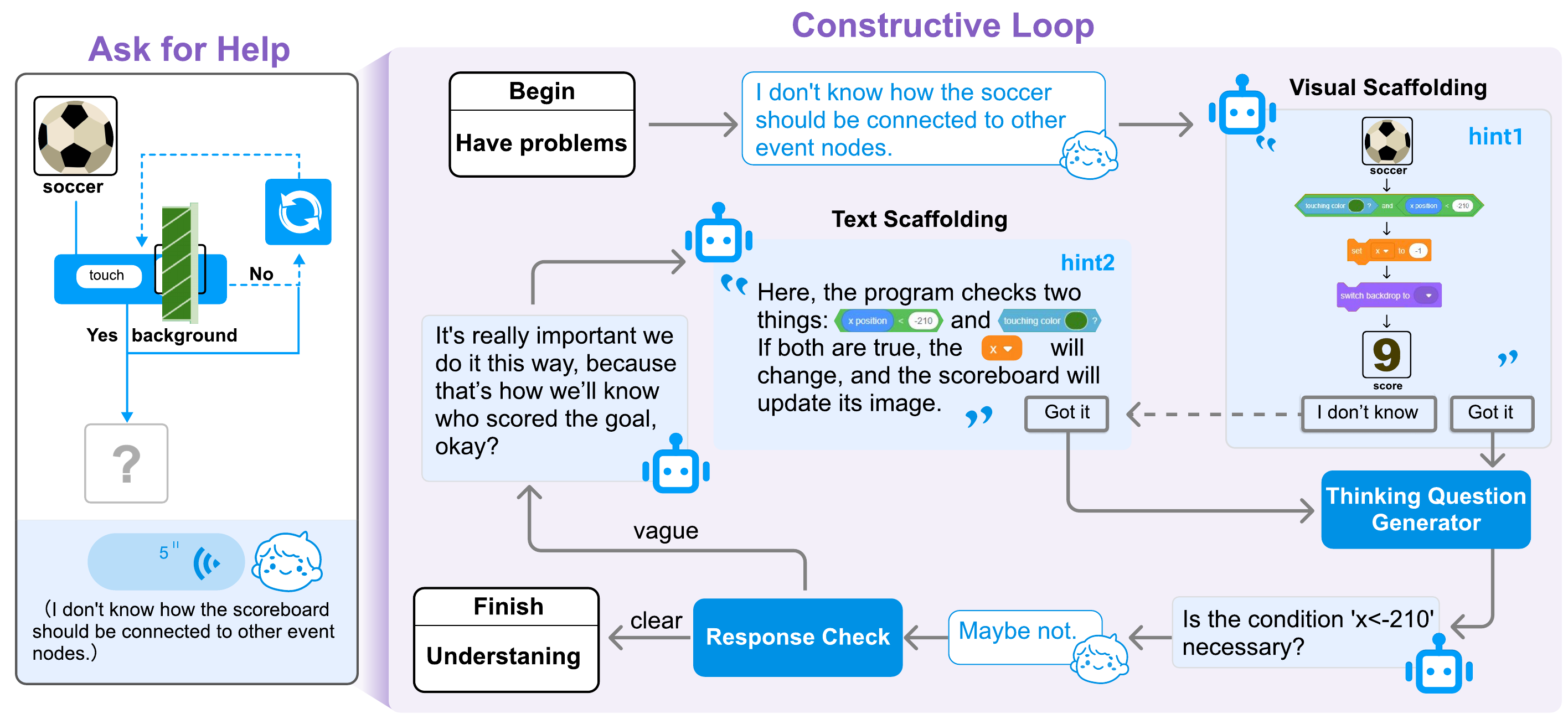}
    \caption{\textbf{Visual and Textual Scaffolding in the Constructive Loop: When a learner poses a question, \textit{CoRemix} provides visual scaffolding to support them. If the learner clicks "Got it," \textit{CoRemix} generates follow-up questions to further assess their understanding. However, if the Response Check of CoRemix is vague or the learner chooses "I don't know," \textit{CoRemix} offers additional textual scaffolding until the learner achieves a clear comprehension.}}
    \vspace{-0.11in}
    \label{fig:vt}
\end{figure*}

When learners encounter difficulties in constructing the visual graph, we employ a question-visual-textual scaffolding framework to guide them in learning computing concepts \textbf{(DG2)}. We identified two main scenarios where children may need scaffolding: (1) when learners are unsure about the relationships between event nodes, and (2) when they are uncertain about how to add CC nodes, such as when they mistakenly place CC nodes between characters. To address the first scenario, the system first determines whether the learner understands the function of the node itself. For example, after a learner drags an event node, CoRemix might respond with, ``What is the scoreboard used for?'' and then provide a visual hint based on the learner’s response (Figure \ref{fig:ui}.(a.2)). For the second scenario, we aim to guide learners to think more critically and build new knowledge. Figure \ref{fig:vt} illustrates the Constructive Loop of CoRemix. In this loop, a learner asks for help connecting the scoreboard with a CC node. CoRemix first responds with visual scaffolding, showing how the elements interact. If the learner remains unclear, the Thinking Question Generator prompts deeper reflection with questions like, ``Is the condition `x<-210' necessary?'' A Response Check evaluates the learner's status. If the response remains vague, textual scaffolding is used to explain the underlying logic with detailed rich text hints, including highlighted character nodes and explanations of the code blocks used in the project.

\subsection{Remixing Projects Based on Visual Graph}
In our formative study, we observed that while learners are motivated to engage in remixing exercises and have creative ideas, the community lacks support for their ideas. The limited asset library does not meet their needs during the remixing phase \cite{chen2024chatscratch}. Therefore, we aim to provide personalized remixing practice by allowing learners to add new nodes and relationships within the visual graph \textbf{(DG3)}. Once learners complete the project understanding phase, they can enter the remixing phase by clicking the ``New Event'' button (Figure \ref{fig:ui}.(a.3). The added events are presented on a canvas, arranged alongside the existing ones, and learners can input their own requirements to select new characters and relationship suggestions generated by the system. For example, a learner might say, ``I want an energy ball in this soccer game.'' CoRemix generates two different energy ball descriptions and images based on the project's context (Figure \ref{fig:ui}.(b.1)). Once the learner selects a node, CoRemix suggests edges to help them build the new graph (Figure \ref{fig:ui}.(b.2)). We encourage learners to use CC nodes to represent computing concepts within events. The created nodes and edges are organized into a graph, breaking down the overall task into manageable sub-tasks, and reducing the programming burden. Learners can then follow the created visual graph to carry out visual programming, producing personally meaningful projects. Additionally, learners can share their computational media with other community members through the visual graph.

To generate new nodes in the visual graph, we first use GPT-4 to convert the user's request into two prompts suitable for a text-to-image generation model. Then, we use the Stable Diffusion model \cite{StableDiffusion2023} to create the node images. An additional GPT-4 model uses the constructed visual graph as a prompt to generate the relations between nodes, including event and computing logic. For the LLM generating image prompts, we used a role-playing strategy and negative prompts \cite{Shanahan2023Role, Kapoor2023NegativePrompts}, instructing it to act as both an educator and a child assistant. This ensures the outputs are educational, child-friendly, and safe for young audiences.

\section{Technical Details}

CoRemix consists of a front-end web page based on the open-source Scratch-GUI framework and a back-end service deployed on Python Flask \footnote{\url{https://github.com/pallets/flask}}. To avoid literacy barriers, we utilized the OpenAI text-to-speech API to deliver voice prompts and transcribe speech \footnote{\url{https://openai.com/research/whisper}}. To protect participants' privacy, we also provided a text input box. To ensure safety, we added a moderation layer to prevent any output containing explicit content, hate speech, harassment, violence, or self-harm \footnote{\url{https://platform.openai.com/docs/api-reference/moderations}}. The final version of the prompts used in our system is shown in the supplementary materials. More technical details will be introduced in this section.

\subsection{Community Resources for RAG}
To enhance the reliability and richness of the visual-textual scaffolding, we use comments and posts as external knowledge sources to provide relevant information to the LLM (Figure \ref{fig:method}). We constructed a Scratch knowledge base and used retrieval-augmented generation techniques to retrieve the knowledge from the community as a context for generating answers. We scraped 23,567 posts and comments from 5,000 popular projects in the Scratch community. We first used GPT-3.5 to extract sentences related to computing concepts, practices, and creative ideas from the comments and replies to posts. We used 200 manually labeled data to fine-tune the GPT-3.5 model for information extraction. For example, someone posted, ``Does anyone have an idea how to make a good inventory for a game?'' and another user replied, ``You can use the List block to store your items in the inventory,'' we extracted sentences containing keywords along with their context as a piece of knowledge. We removed sentences longer than 400 words or shorter than 5 words, as these were mostly extraction errors. We then used semantic embeddings to merge semantically similar sentences. After further manual review, we finalized a set of 3,528 sentences. In the understanding and remixing phases, if a user poses a question \(q\), we first retrieve the top three most relevant sentences from the knowledge base \(D = \{d_1, d_2, d_3, \ldots, d_n\}\), and use these sentences as context \(R(q, D) = \{d_{q1}, d_{q2}, d_{q3}\}\) for the generative language model to produce an answer \(\text{LLM}(q, R(q, D))\).

\subsection{Project Analysis}
\begin{figure*}[htp]
    \centering
    \includegraphics[width=1\textwidth]{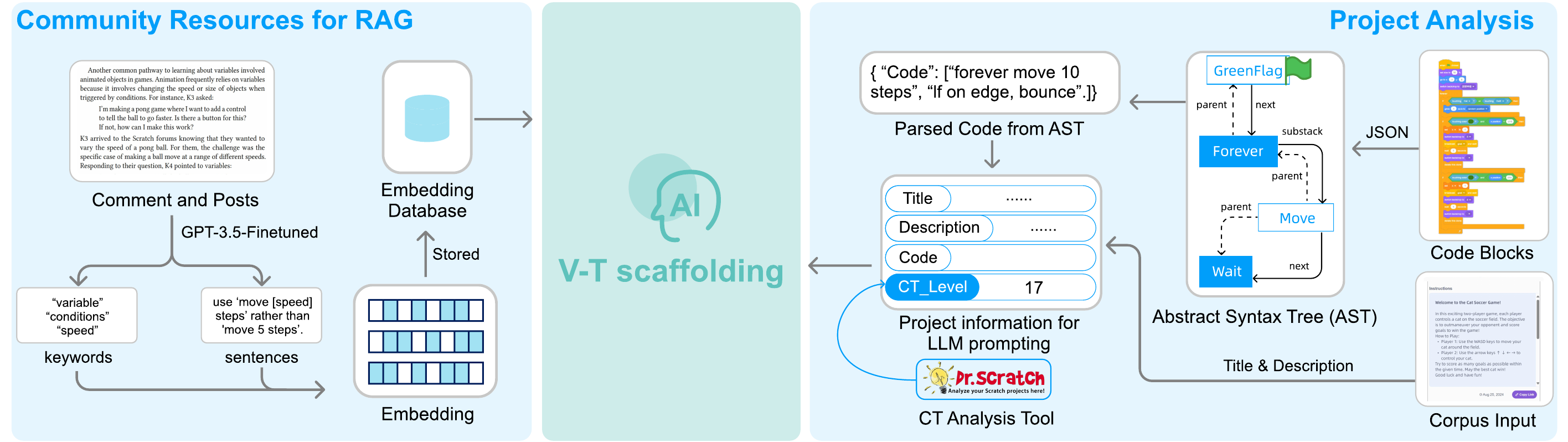}
    \caption{\textbf{Workflow of community resource-driven retrieval-augmented generation and project analysis.}}
    \vspace{-0.11in}
    \label{fig:method}
\end{figure*}

A Scratch project consists of code blocks, images, and audio assets. To enable the LLM to understand the multimodal information of projects and provide more reliable answers when learners construct visual graphs, we employed a novel method for parsing Scratch projects. This method parses the ``project.json'' file by generating an abstract syntax tree (AST) based on defined grammar rules. Because Scratch allows flexible naming for sprites, blocks, and variables, we avoid ambiguities by using basic JSON syntax for parsing. As shown in Figure \ref{fig:method}, the AST's leaf nodes correspond to words in the input file, and as we traverse the AST, we record block information and relationships to create a block tree for LLM prompting. In the block tree, each node represents a block with attributes such as name, input, and field information. The relationships between nodes include parent, next, and substack. For example, the ``green flag'' block is the parent of the ``forever'' block, which also has ``next'' and ``substack'' blocks. The block tree is then traversed using depth-first search to analyze the project according to the CT skills criteria from Dr.Scratch \cite{moreno2015dr}, which evaluates CT scores and is used here to measure a project's complexity. we use the parsed project context as the GPT-4's memory to enhance its understanding of community projects (Figure \ref{fig:method}).

\subsection{Prompting Pipeline for Visual-Textual Dialogues}
Inspired by \cite{jin2024teach}, CoRemix adopts a constructive loop that uses multi-stage visual and textual scaffolding to make the dialogue process dynamic and appropriately poses questions to encourage learners to engage in deeper thinking (Figure \ref{fig:vt}). We use a chain-of-thought \cite{wu2022ai} to have GPT-4 first generate a summary answer, followed by a graph that includes all control dependencies of the code block  \(\text{generated\_block} = \{\text{block\_1}, \text{block\_2}, \ldots, \text{block\_n}\}\). CoRemix provides feedback by visualizing the generated blocks using the Scratch block's unique ID to create the graph. The generated graph is rendered with Scratch blocks as nodes, each showing arrows indicating its associated targets. Additionally, we introduce a Thinking Question Generator module that uses GPT-4 and RAG \cite{fan2024survey} to produce thought-provoking questions related to the current conversation. The Response Check assesses each learner's answer in the loop and, based on the constructive learner inquiry protocol \cite{shahriar2021can}, determines whether to provide detailed textual scaffolding. The textual scaffolding is generated with retrieval augmented GPT-4 that incorporates valuable community resources.

\subsection{Technical Evaluation}
We selected 10 projects with the same level of difficulty in the CT score dimension from the collected projects and used the project analysis module to extract the code blocks and corpus as the project context. For fairness, we used project descriptions provided by beginners as the baseline project context. These two project contexts were then used to create prompts in the same format to prompt GPT-4, referred to as project-based LLM and novice-based LLM. We tasked both the project-based LLM and novice-based LLM with generating three items to reflect their understanding of the project, including relationships, project flow, and the use of computing concepts. Two education experts with over five years of experience were invited to evaluate the generated content using a 7-point Likert scale (1 - Strongly Disagree, 7 - Strongly Agree). Table \ref{tab:llm1} shows the results, where the project-based LLM outperformed the baseline on all dimensions. This indicates that the project analysis module, by integrating code blocks and corpus information, enables the LLM to achieve a more comprehensive understanding of the project, thereby improving the quality of dialogue between learners and the system.

\begin{table}[ht]
\centering
\caption{\textbf{Comparison of LLM Performance in Understanding Scratch Projects.}}
\label{tab:llm1}
\footnotesize
\vspace{-0.11in}
\begin{tabular}{lccc}
\toprule
\textbf{} & \textbf{Relationships} & \textbf{Project Flow} & \textbf{Use of computing Concepts} \\
\midrule
\textbf{Novice-based}   & ${5.4\pm1.11}$  & ${6.0\pm0.89}$  & ${5.8\pm0.98}$ \\
\textbf{Project-based}  & ${\textbf{5.7}\pm0.78}$  & ${\textbf{6.3}\pm0.79}$  & ${\textbf{5.9}\pm0.83}$ \\
\bottomrule
\end{tabular}
\end{table}

We also evaluated the impact of the RAG module on the quality of the conversational agent's responses. Two education experts wrote 10 questions based on the 10 previously mentioned projects. We then used a vanilla LLM as the baseline and RA-LLM to generate answers to these 10 questions. The experts rated the answers across four dimensions: Relevance, Richness of Content, Fluency, and Educational Value, using a 7-point Likert scale (1 - Strongly Disagree, 7 - Strongly Agree). Table \ref{tab:llm2} shows the results, where RA-LLM outperformed the baseline regarding Richness of Content and Educational Value. This indicates that the community resource knowledge base can help the LLM provide a broader range of knowledge and more accurately address the learners' questions.

\begin{table}[ht]
\centering
\caption{\textbf{Comparison of LLM Performance in Response Quality}}
\label{tab:llm2}
\footnotesize
\vspace{-0.11in}
\begin{tabular}{lcccc}
\toprule
\textbf{} & \textbf{Relevance} & \textbf{Richness of Content} & \textbf{Fluency} & \textbf{Educational Value} \\
\midrule
\textbf{Vanilla LLM}   & ${\textbf{5.2}\pm1.87}$  & ${4.7\pm1.18}$  & ${5.3\pm1.1}$ & ${4.2\pm1.07}$ \\
\textbf{Retrieval-augmented LLM}  & ${4.8\pm0.60}$  & ${\textbf{6.1}\pm0.83}$  & ${\textbf{5.7}\pm0.9}$ & ${\textbf{5.2}\pm0.97}$ \\
\bottomrule
\end{tabular}
\end{table}
\section{Evaluation}

To evaluate the effectiveness of CoRemix in project understanding and remixing activity, we conducted a within-subjects study, comparing the CoRemix condition to the Scratch community webpage as the baseline. The study was structured around three primary research questions:

\begin{itemize}
    \item \textbf{RQ1.} How does CoRemix enhance learners' understanding of community projects through visual graphs?
    \item \textbf{RQ2.} How does CoRemix help learners develop computing concepts through visual graph building and visual-textual scaffolding?
    \item \textbf{RQ3.} How does CoRemix facilitate learners' remixing practices in the community by supporting the creation of new nodes and edges within the visual graph?
    
\end{itemize}

\subsection{Participants}

We recruited 16 participants (P1-P16; 6 female, 10 male; aged 9 to 12, M = 10.06, SD = 1.02) from two programming education institutions. Each participant took part in both experimental conditions, exploring two community projects—a soccer game and a racing game—in separate sessions. According to our preliminary survey, all participants were elementary school students and novice learners in the Scratch Community, with less than one year of experience in programming.

\subsection{Experimental Design}

\subsubsection{Design of Learning Tasks.} In a within-subjects study, each participant completed two theme-based learning tasks: one using CoRemix and the other using a baseline tool. Based on previous research \cite{park2017telling} and recommendations from educators, we selected two game projects from the community: a soccer game (Theme A) and a racing animation (Theme B). According to the CT criteria from Dr. Scratch, both projects were of equal difficulty. In each task, participants were required to understand the computing concepts embedded in the project’s events and engage in remixing practices, such as creating new characters or programming events to add their creative elements. At the end of each task, participants were asked to provide a comprehensive description of the project focusing on four dimensions: the project’s goal, key events, project detail, and relationships between events, to assess their understanding of the project. They also completed a subjective questionnaire to evaluate their user experience and cognitive load with the system, as well as an objective test with 14 multiple-choice questions (MCQs) to assess their grasp of the computing concepts in the project.

\subsubsection{Baseline: \textit{The Scratch Community Webpage}} To evaluate the effectiveness of CoRemix in supporting novice learners during informal learning, we selected the Scratch community webpage as a baseline for two key reasons. First, Scratch is recognized in educational research as a representative platform where novice learners informally explore and learn programming projects \cite{romero2021scratch,dasgupta2017scratch,deiner2023automated}. Second, since the community resource knowledge base in the CoRemix system is derived from Scratch, using it as a baseline ensures consistency in the learning materials. Participants in the baseline condition could explore resources within the Scratch community, such as user-created comments, posts, and projects.

\subsection{Procedure}
\begin{figure*}[htp]
    \centering
    \includegraphics[width=0.95\linewidth]{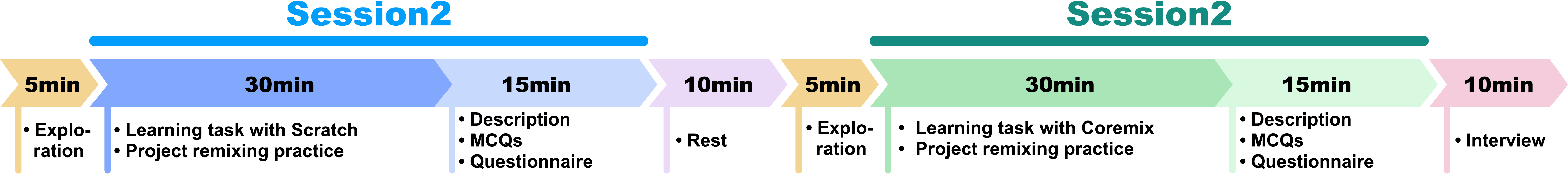}
    \caption{\textbf{The process of participants engaging in the within-subjects study.}}
    \label{flow}
    \vspace{-0.11in}
\end{figure*}

As shown in Figure \ref{flow}, before each session, participants were given 5 minutes to learn about the task and explore the system freely, with an opportunity to ask questions to the researchers. Each session lasted around 45 minutes: 20 minutes for completing a theme-based learning task and 10 minutes for remixing the project and submitting it. 15 minutes for describing the project, as well as to complete the questionnaires and MCQs. Participants had a 10-minute break between sessions. After both sessions, we conducted semi-structured interviews to gather insights into their experiences. Researchers only assisted with technical issues during the tasks. We also collected video data through screen recordings and side-angle footage. Finally, each participant received \$20 in compensation.

\subsection{Measures}
We focus on learners' understanding and learning of community projects, the creative support they experience during remixing activities, and their sense of engagement and immersion throughout the community learning process. We summarized our metrics and employed the Bonferroni correction for all statistical tests with the questionnaires to avoid potential multiple comparison problems.

\textbf{RQ1. Understanding of community projects.}
Based on existing research in block-based programming education \cite{fagerlund2021computational,saez2016visual}, we developed four metrics to evaluate learners' project descriptions as a reflection of their understanding of the project: (1) Project Goal: assesses the participant's ability to analyze the project's output and objectives; (2) Key Event: measures the ability to abstract and identify key events, reflecting the understanding of core programming logic; (3) Project Detail: evaluates the skill in breaking down specific events, including code functionality and parameters; (4) Logical Relationship: assesses understanding of event relationships, such as parallelism and sequence, and the coherence of the programming logic. To conduct this assessment, three experienced educators independently rated the participants' descriptions using a 5-point Likert scale (1 - Not understood at all, 5 - Fully understood). Expert evaluations demonstrated a high level of inter-rater reliability, with a significant intraclass correlation coefficient (ICC) of 0.82 (p < .001).

\textbf{RQ2. Learning of computing concepts.}
Based on computational thinking scales \cite{korkmaz2017validity, tang2020assessing}, we developed a multiple-choice questionnaire to assess participants' ability to define, recall, and transfer computing concepts. The questionnaire consists of seven dimensions: abstraction, parallelism, logic, synchronization, flow control, interactivity, and data representation, with two questions per dimension. The questions are generated based on specific programming projects. Each question has four answer choices, with 1 point awarded for a correct answer and 0 points for an incorrect one. For example, to assess participants' understanding of abstraction in a soccer game project, one question is: ``What is the role of \textit{variable} in the code blocks of this project?'' According to this scoring system, the total score is calculated by summing all item scores, ranging from 0 to 14.

\textbf{RQ3. Creativity support of remixing practice.}
We analyzed learners' remixing performance through two metrics related to the visual graph: the number of extended nodes and extended edges. To measure learners' perceived creative support, we adapted a 4-item questionnaire based on \cite{cherry2014quantifying}, including enjoyment, exploration, expressiveness, and immersion, with each item using a 5-point Likert scale. To measure learners' perceived cognitive load under different conditions, we used Morrison et al.'s questionnaire \cite{morrison2014measuring} designed for CS learning. We adapted the questionnaire to a 5-point Likert scale and modified the questions to fit younger learners' cognitive levels.

\subsection{Results}

In this section, we systematically answer the three research questions based on evidence from the data. 

\textbf{[RQ1] CoRemix enhances learners' ability to decompose projects into parts, thereby improving their understanding of key events and relationships within the project.}

We found that learners in the CoRemix condition scored higher on the overall project understanding compared to the baseline across all dimensions (Total Score: \textit{Baseline} = \(11.63 \pm 0.96\), \textit{CoRemix} = \(14 \pm 0.55\))). As shown in Table \ref{Table:understanding}, learners showed significant effects in identifying key events (\textit{Baseline} = \(2.63 \pm 0.74\), \textit{CoRemix} = \(3.21 \pm 0.47\), two-tailed t-test, \(p < 0.05^{*}\)) and project details (\textit{Baseline} = \(2.85 \pm 0.89\), \textit{CoRemix} = \(3.65 \pm 0.59\), two-tailed t-test, \(p < 0.05^{*})\). This may be due to the visual abstraction provided by the visual graph, which enhances learners' ability to decompose projects by breaking down programming events into multiple canvases and abstracting programming logic into sequential connections between CS nodes and character nodes. Learners also improved in programming logic (\textit{Baseline} = \(3.21 \pm 0.77\), \textit{CoRemix} = \(3.85 \pm 0.53\), two-tailed t-test, \(p < 0.01^{**}\)). P8 mentioned, ``Before using it (\textit{CoRemix}), I could only find some simple projects to learn, but after building the visual graph, I could understand complex projects more easily.'' Under the baseline condition, we found that learners' understanding of the project often lacked abstraction and a deeper grasp of relationships. For example, they tended to focus more on the functions of individual characters but missed the interactions between them. We did not observe significant differences in project goals, which might be because the goals of Scratch projects are often straightforward.

\begin{table*}[htp]
\centering
\footnotesize
\caption{\textbf{Expert Scores on Project Understanding for the Baseline and CoRemix.}}
\vspace{-0.11in}
\label{Table:understanding}
\begin{tabularx}{\textwidth}{p{0.2\textwidth} X X X X X X}
\toprule
\multirow{2}{*}{Metric} & \multicolumn{2}{c}{Baseline} & \multicolumn{2}{c}{CoRemix} & \multicolumn{2}{c}{Paired-t test} \\
\cmidrule(r){2-3} \cmidrule(r){4-5} \cmidrule(r){6-7}
& Mean & SD & Mean & SD & t & p \\
\midrule
Project Goal & 2.94 & 1.35 & 3.29 & 0.36 & -1.10 & 0.288   \\
Key Event    & 2.63 & 0.74 & 3.21 &  0.47 & -2.31 & 0.035*  \\
Project Detail     &  2.85 & 0.89 &  3.65 &  0.59 & -2.84 & 0.012*  \\
Logical Relationship & 3.21 & 0.77 & 3.85 &  0.53 & -3.14 & 0.006** \\
\bottomrule
\end{tabularx}
 \end{table*}

\textbf{[RQ2] Visual-textual scaffolding promotes learners' understanding of computing concepts through visual graph construction.}

We also analyzed the impact of CoRemix on learners' learning outcomes compared to the baseline. As shown in Table \ref{Table:ct}, we observed significant differences in four dimensions (abstraction, parallelism, synchronization, and data). Participants' feedback indicated that the visual-textual scaffolding significantly contributed to their understanding of computing concepts. Learner feedback indicated that the process of constructing the visual graph contributed to their understanding of abstraction (\textit{Baseline}: Mean = 0.69, SD = 0.94 / \textit{CoRemix}: Mean = 1.62, SD = 0.50 / \(t = -4.85, p<0.01^{**}\)) and parallel (\textit{Baseline}: Mean = 0.81, SD = 0.75 / \textit{CoRemix}: Mean = 1.43, SD = 0.51 / \(t = -2.61, p<0.05^{*}\)). We found that the visual-textual scaffolding significantly contributed to their understanding of synchronization (\textit{Baseline}: Mean = 0.94, SD = 0.68 / \textit{CoRemix}: Mean = 1.62, SD = 0.71 / \(t = -2.55, p<0.05^{*}\)) and data representation (\textit{Baseline}: Mean = 0.93, SD = 0.85 / \textit{CoRemix}: Mean = 1.687, SD = 0.47 / \(t = -3.39, p<0.01^{**}\)).  Additionally, we observed that the interaction with CoRemix, while receiving visual and textual answers, motivated learners to be more active in their learning. When CoRemix provided an overview of the code graph or textual answers with code blocks, some learners frequently revisited the original project or clicked on the CC nodes in the visual graph while thinking of answers (P2, P5, P6, P10). Meanwhile, other learners showed excitement when they received the correct answers (P1, P4, P7, P8). P7 mentioned, ``I can focus better to get the correct answer. When I get the answer, I think of ways to verify it.'' This finding aligns with previous research, which indicates that meaningful learning occurs when learners actively think about the presented information rather than passively receiving it \cite{chi2014icap}. Additionally, learners generally found CoRemix's guidance easier to understand for the computing concepts in the projects compared to the baseline condition. P2, P5, P6, and P10 mentioned that they preferred CoRemix because its explanations were easier to understand.

\begin{table*}[htp]
\centering
\footnotesize
\caption{\textbf{Comparison of the number of correct MCQs on computing concepts between the Baseline and Coremix.}}
\vspace{-0.11in}
\label{Table:ct}
\begin{tabularx}{\textwidth}{p{0.15\textwidth} X X X X X X}
\toprule
\multirow{2}{*}{Metric} & \multicolumn{2}{c}{Baseline} & \multicolumn{2}{c}{CoRemix} & \multicolumn{2}{c}{Paired-t test} \\
\cmidrule(r){2-3} \cmidrule(r){4-5} \cmidrule(r){6-7}
& Mean & SD & Mean & SD & t & p \\
\midrule
Abstraction     & 0.69 & 0.94 & 1.62 & 0.50 & -4.85 & 0.002** \\
Parallelism     & 0.81 & 0.75 & 1.43 & 0.51 & -2.61 & 0.020*  \\
Logical         & 1.06 & 0.77 & 1.25 & 0.44 & -1.71 & 0.333  \\
Synchronization & 0.94 & 0.68 & 1.62 & 0.71 & -2.55 & 0.016* \\
Flow Control    & 1.19 & 0.83 & 1.43 & 0.87 & -1.69 & 0.331   \\
Interactivity   & 1.32 & 0.91 & 1.56 & 0.51 & -1.30 & 0.117 \\
Data            & 0.93 & 0.85 & 1.68 & 0.47 & -3.39 & 0.009**  \\
\midrule
Total Score & 6.94 & 2.40 & 10.59 & 1.13 & -6.13 & 0.001** \\
\bottomrule
\end{tabularx}
\end{table*}

\begin{figure*}[htp]
    \centering
    \includegraphics[width=1\textwidth]{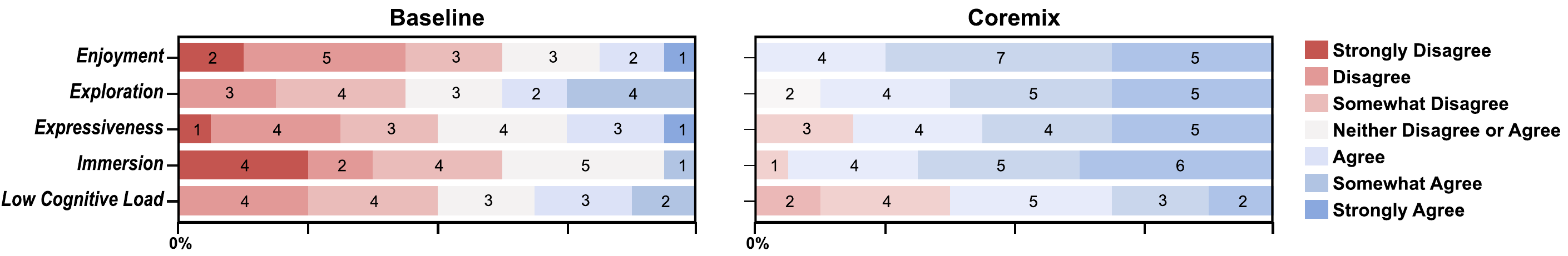}
    \caption{\textbf{Distribution of user ratings on the Baseline and CoRemix.}}
    \label{rating}
    \vspace{-0.11in}
\end{figure*}

\textbf{[RQ3] CoRemix increases learners' engagement of computing participation and improves their creativity of remixing practice by providing support for node and edge construction.}
First, we evaluated the effectiveness of supporting remixing activities within the visual graph, focusing on how extensively learners adopted the provided node and relationship suggestions. The average number of extended nodes was 3.59 with a standard deviation of 1.25, while the average number of extended edges was 6.79 with a standard deviation of 3.45, indicating that learners made considerable modifications to the original projects. Notably, learners used node and relationship suggestions an average of 4.84 times, with a standard deviation of 0.74, suggesting that the visual graph format better stimulated their creativity and encouraged more active involvement in the project development process. Figure \ref{rating} compares the distribution of scores from the creativity support and cognitive load questionnaires under the baseline and CoRemix conditions. In every question, CoRemix outperformed the baseline, indicating its usefulness in supporting creativity (enjoyment, exploration, expressiveness, and immersion) and managing cognitive load. Learners felt that CoRemix effectively facilitated creative expression (\textit{Baseline}: Mean = 3.50, SD = 1.40 / \textit{CoRemix}: Mean = 5.93, SD = 1.08 / \(t = -3.75, p < 0.01^{**}\)) and exploration (\textit{Baseline}: Mean = 4.00, SD = 1.45 / \textit{CoRemix}: Mean = 5.81, SD = 1.01 / \(t = -3.95, p<0.01^{**}\)), providing a more enjoyable (\textit{Baseline}: Mean = 3.13, SD = 1.57 / \textit{CoRemix}: Mean = 6.06, SD = 0.74 / \(t = -6.52, p<0.01^{**}\)) and immersive experience (\textit{Baseline}: Mean = 2.87, SD = 1.40 / \textit{CoRemix}: Mean = 5.93, SD = 1.08 / \(t = -6.66, p<0.01^{**}\)) (Table \ref{Table:rating}). This improvement may be attributed to the timely support provided during remixing activities, which stimulated creativity and encouraged active learning. Learners also expressed more pride in the programming projects they created and were more willing to share their artifacts within the community (P1, P2, P5, P7, P9). We observed no statistically significant difference in cognitive load (\textit{Baseline}: Mean = 3.68, SD = 1.35 / \textit{CoRemix}: Mean = 4.56, SD = 1.61 / \(t = -1.60, p = 0.119\)). However, the CoRemix condition exhibited a lower average cognitive load. This result suggests that introducing visual graphs to support understanding and using visual-textual scaffolds to enhance feedback not only avoided increasing cognitive load but also helped maintain a manageable load during extended learning tasks.

\begin{table*}[htp]
\centering
\footnotesize
\caption{\textbf{Comparison of the Baseline and CoRemix Across User Ratings.}}
\vspace{-0.11in}
\label{Table:rating}
\begin{tabularx}{\textwidth}{p{0.15\textwidth} X X X X X X}
\toprule
\multirow{2}{*}{Metric} & \multicolumn{2}{c}{Baseline} & \multicolumn{2}{c}{CoRemix} & \multicolumn{2}{c}{Paired-t test} \\
\cmidrule(r){2-3} \cmidrule(r){4-5} \cmidrule(r){6-7}
& Mean & SD & Mean & SD & t & p \\
\midrule
Enjoyment     & 3.13 & 1.57 & 6.06 & 0.74 & -6.52 & 0.000** \\
Exploration     & 4.00 & 1.45 & 5.81 & 1.01 & -3.95 & 0.004**  \\
Expressiveness   & 3.50 & 1.50 & 5.50 & 1.41 & -3.75 & 0.007**  \\
Immersion       & 2.87 & 1.40 & 5.93 & 1.08 & -6.66 & 0.000** \\
Cognitive Load    & 3.68 & 1.35 & 4.56 & 1.61 & -1.60 & 0.119   \\
\bottomrule
\end{tabularx}
\end{table*}
\section{Discussion}
In this study, we propose a community-driven approach to learning CT concepts and practices through interaction with an LLM-powered visual graph. Our method focuses on abstracting community projects into a visual graph, supported by visual-textual scaffolding within a constructive loop to enhance project understanding and facilitate computational learning. We found that novices using CoRemix demonstrated a better understanding of computing concepts and more effective engagement in computing practices compared to the community interface baseline. Learners attributed these benefits to our visual graph and visual-textual scaffolds. We built a community resource knowledge base and used RA-LLM to enhance the reliability of responses. Our findings encourage further research into generative agents that support learning in programming communities. Our approach complements previous studies on programming education tools, such as ChatScratch \cite{chen2024chatscratch}, Sara \cite{winkler2020sara}, and Codeaid \cite{kazemitabaar2024codeaid}. In this section, we discuss design recommendations, advantages, limitations, and future directions for GAI-driven tools that support informal community learning.

\subsection{Design Considerations for Constructing Visual Graphs}
Some believe that informal learning systems should provide "wide walls," supporting learners’ ability to creatively construct their own meaning through multiple possibilities and pathways \cite{cheng2022many}. In the context of Scratch, the unstructured nature of the Scratch online community has led to an over-representation of certain knowledge applications, thereby "narrowing" these walls \cite{cheng2022interest}. Our work utilizes visual graphs that integrate computing concepts as starting points for understanding projects. While creating these graphs, we provide iterative visual and textual scaffolding, following a constructive loop, to guide learners toward more active engagement with project examples and to inspire remixing activities. CoRemix employs this scaffolding to encourage learners to first put in effort, followed by support and feedback. For instance, during the visual graph construction process, CoRemix continuously guides learners with questions until they reach satisfactory answers. However, this approach also increased learners' cognitive load, although we did not observe significant differences in cognitive load between CoRemix and the baseline. As learners developed their computing concepts, three CoRemix users (P4, P12, P15) pointed out that the visual graph could not fully represent project details. Specifically, when the conversational agent provided incorrect answers to their questions, they were unsure how to correct the graph (P12, P15). This suggests the need for additional knowledge acquisition channels to supplement the conversational agent. Socratic questioning \cite{knezic2010socratic, jin2024teach}, tailored to specific knowledge domains, could help learners grasp unfamiliar concepts by posing thought-provoking questions at appropriate moments without disrupting the intended learning process.

In summary, to encourage learners to reflect on their artifacts and engage in deeper thinking, future research on programming learning support tools could adopt a design similar to CoRemix, combining question-driven guidance with hands-on practice. However, our findings suggest a trade-off between these two strategies. To foster knowledge construction, the system should intervene with responses that include thought-provoking questions. Conversely, to make the system more engaging and enhance practical skills, learners should be given more freedom to engage in hands-on activities. Future research could empirically explore this trade-off and how learners balance these strategies when they have direct control over the use of community resources.

\subsection{Generative AI for Expanding Opportunities for Community Participation}
In this study, we propose a promising approach that uses visual graphs equipped with a generative AI-powered conversational agent as a good starting point for informal learning. Our method of fine-tuning LLMs to extract useful knowledge points from the community and constructing a knowledge base can transfer to similar methods in other fields, such as learning UI layout design \cite{oppenlaender2020crowdui, lovell2023lilytiny} or video editing \cite{burgess2020agency}. Researchers can follow the workflow we detailed in Section 5 to generate learning materials. They can then explore using retrieval-augmented generation \cite{wiratunga2024cbr} to provide specific examples and employ prompt engineering to generate more educationally meaningful responses. We also suggest that researchers adopt more open-ended questions to stimulate active thinking in learners, such as using mode switching and learning-by-teaching approaches \cite{jin2024teach}.

Programming is a crucial tool for teaching, learning, and assessing students' computational thinking in K-12 classrooms, but it is not the only way to foster CT skills. Alternative approaches, such as articulating precise pseudo-code, creating flowcharts, and developing step-by-step processes, can also encourage learners to computing practice \cite{gunbatar2019computational}. This study explores the use of generative AI to support informal learning in the Scratch programming community. In our experiment, CoRemix, leveraging a community knowledge base and retrieval-based generation, left a strong impression on educators. Compared to aimless exploration within the community, these AI-generated responses made learners feel as if they were engaging in peer learning. However, learners noted that the generated content sometimes contained errors, a phenomenon known as a hallucination in LLMs \cite{liu2024exploring}. While hallucinations are currently unavoidable, we observed that learners tended to reflect critically when faced with uncertain answers and were able to identify the model’s hallucinations. To some extent, this can be attributed to the deeper understanding participants developed while co-creating the visual graph with CoRemix.

Moreover, it is important to recognize the benefits of generative AI in inspiring remixing activities. Several studies have shown that when learners' interests and identities are underrepresented in computational learning communities, they may feel excluded or marginalized \cite{lovell2023lilytiny, richard2016blind, ford2016paradise}. For instance, although many girls use Scratch, there is evidence that girls generally show less interest in game creation compared to boys \cite{funke2017gender, hsu2014gender}. To address this challenge, community researchers might consider using generative AI to help users create projects aligned with their interests. With advanced image generation technologies, researchers could also track users' activities within the community to create interest maps, which can be used to generate more personalized visual assets and code that matches the users' current programming abilities. Our findings offer a potential pathway for guiding such work.

\subsection{Limitations and Future Work}
Our work has several limitations that need to be addressed in future research. First, although our RAG-based conversational agent performed well on human evaluation metrics, it still faces challenges with response relevance and handling complex questions. Future work could improve the response's performance by expanding the size of the knowledge base and applying the latest LLMs. Second, while we evaluated the impact of the tool on learners' learning outcomes and user experience in our experiments, we did not conduct long-term studies. Therefore, our empirical results cannot rule out the novelty effect of using CoRemix for the first time. Lastly, our participants were novice users with little to no experience using the Scratch community. It would be interesting to study how regular community participants use CoRemix and online programming resources for learning. For future work, we suggest measuring the long-term learning experiences and outcomes of users with different backgrounds.

\section{Conclusion}

In this study, we introduced CoRemix, an interactive generative AI-supported system designed to facilitate informal visual programming learning in interest-driven communities for beginners aged 9-12. CoRemix guides learners in constructing visual graphs to understand projects and supports computing practice by allowing them to create nodes and establish edges within the visual graph. To enhance its effectiveness, we integrated community knowledge-based RA-LLMs, providing visual-textual scaffolding to aid in the comprehension of computing concepts and knowledge building. Additionally, generative AI was utilized to support creative remixing activities. A study with 16 novice learners aged 9-12 demonstrated the effectiveness of CoRemix in enhancing project understanding, developing computing concepts, and improving learning experiences. We also offer insights into expanding this approach to other online communities, and we hope this work encourages further exploration of community resources for building intelligent informal learning support systems.
\bibliographystyle{ACM-Reference-Format}
\bibliography{sections/Reference}


\begin{thebibliography}{79}


\ifx \showCODEN    \undefined \def \showCODEN     #1{\unskip}     \fi
\ifx \showDOI      \undefined \def \showDOI       #1{#1}\fi
\ifx \showISBNx    \undefined \def \showISBNx     #1{\unskip}     \fi
\ifx \showISBNxiii \undefined \def \showISBNxiii  #1{\unskip}     \fi
\ifx \showISSN     \undefined \def \showISSN      #1{\unskip}     \fi
\ifx \showLCCN     \undefined \def \showLCCN      #1{\unskip}     \fi
\ifx \shownote     \undefined \def \shownote      #1{#1}          \fi
\ifx \showarticletitle \undefined \def \showarticletitle #1{#1}   \fi
\ifx \showURL      \undefined \def \showURL       {\relax}        \fi
\providecommand\bibfield[2]{#2}
\providecommand\bibinfo[2]{#2}
\providecommand\natexlab[1]{#1}
\providecommand\showeprint[2][]{arXiv:#2}

\bibitem[Barrot(2022)]%
        {barrot2022social}
\bibfield{author}{\bibinfo{person}{Jessie~S Barrot}.} \bibinfo{year}{2022}\natexlab{}.
\newblock \showarticletitle{Social media as a language learning environment: a systematic review of the literature (2008-2019)}.
\newblock \bibinfo{journal}{\emph{Computer assisted language learning}} \bibinfo{volume}{35}, \bibinfo{number}{9} (\bibinfo{year}{2022}), \bibinfo{pages}{2534--2562}.
\newblock


\bibitem[Bruckman(1998)]%
        {bruckman1998community}
\bibfield{author}{\bibinfo{person}{Amy Bruckman}.} \bibinfo{year}{1998}\natexlab{}.
\newblock \showarticletitle{Community support for constructionist learning}.
\newblock \bibinfo{journal}{\emph{Computer Supported Cooperative Work (CSCW)}}  \bibinfo{volume}{7} (\bibinfo{year}{1998}), \bibinfo{pages}{47--86}.
\newblock


\bibitem[Burgess et~al\mbox{.}(2020)]%
        {burgess2020agency}
\bibfield{author}{\bibinfo{person}{Jean Burgess}, \bibinfo{person}{Joshua Green}, {and} \bibinfo{person}{Gala Rebane}.} \bibinfo{year}{2020}\natexlab{}.
\newblock \showarticletitle{Agency and controversy in the YouTube community}.
\newblock \bibinfo{journal}{\emph{Handbuch soziale praktiken und digitale alltagswelten}} (\bibinfo{year}{2020}), \bibinfo{pages}{105--116}.
\newblock


\bibitem[Burke et~al\mbox{.}(2016)]%
        {burke2016computational}
\bibfield{author}{\bibinfo{person}{Quinn Burke}, \bibinfo{person}{W~Ian O'Byrne}, {and} \bibinfo{person}{Yasmin~B Kafai}.} \bibinfo{year}{2016}\natexlab{}.
\newblock \showarticletitle{Computational participation: Understanding coding as an extension of literacy instruction}.
\newblock \bibinfo{journal}{\emph{Journal of adolescent \& adult literacy}} \bibinfo{volume}{59}, \bibinfo{number}{4} (\bibinfo{year}{2016}), \bibinfo{pages}{371--375}.
\newblock


\bibitem[Campbell et~al\mbox{.}(2016)]%
        {campbell2016thousands}
\bibfield{author}{\bibinfo{person}{Julie Campbell}, \bibinfo{person}{Cecilia Aragon}, \bibinfo{person}{Katie Davis}, \bibinfo{person}{Sarah Evans}, \bibinfo{person}{Abigail Evans}, {and} \bibinfo{person}{David Randall}.} \bibinfo{year}{2016}\natexlab{}.
\newblock \showarticletitle{Thousands of positive reviews: Distributed mentoring in online fan communities}. In \bibinfo{booktitle}{\emph{Proceedings of the 19th ACM Conference on Computer-Supported Cooperative Work \& Social Computing}}. \bibinfo{pages}{691--704}.
\newblock


\bibitem[Chaiklin et~al\mbox{.}(2003)]%
        {chaiklin2003zone}
\bibfield{author}{\bibinfo{person}{Seth Chaiklin} {et~al\mbox{.}}} \bibinfo{year}{2003}\natexlab{}.
\newblock \showarticletitle{The zone of proximal development in Vygotsky’s analysis of learning and instruction}.
\newblock \bibinfo{journal}{\emph{Vygotsky’s educational theory in cultural context}} \bibinfo{volume}{1}, \bibinfo{number}{2} (\bibinfo{year}{2003}), \bibinfo{pages}{39--64}.
\newblock


\bibitem[Chen et~al\mbox{.}(2024)]%
        {chen2024chatscratch}
\bibfield{author}{\bibinfo{person}{Liuqing Chen}, \bibinfo{person}{Shuhong Xiao}, \bibinfo{person}{Yunnong Chen}, \bibinfo{person}{Yaxuan Song}, \bibinfo{person}{Ruoyu Wu}, {and} \bibinfo{person}{Lingyun Sun}.} \bibinfo{year}{2024}\natexlab{}.
\newblock \showarticletitle{ChatScratch: An AI-Augmented System Toward Autonomous Visual Programming Learning for Children Aged 6-12}. In \bibinfo{booktitle}{\emph{Proceedings of the CHI Conference on Human Factors in Computing Systems}}. \bibinfo{pages}{1--19}.
\newblock


\bibitem[Chen and Hwang(2020)]%
        {chen2020effects}
\bibfield{author}{\bibinfo{person}{Mei-Rong~Alice Chen} {and} \bibinfo{person}{Gwo-Jen Hwang}.} \bibinfo{year}{2020}\natexlab{}.
\newblock \showarticletitle{Effects of a concept mapping-based flipped learning approach on EFL students’ English speaking performance, critical thinking awareness and speaking anxiety}.
\newblock \bibinfo{journal}{\emph{British Journal of Educational Technology}} \bibinfo{volume}{51}, \bibinfo{number}{3} (\bibinfo{year}{2020}), \bibinfo{pages}{817--834}.
\newblock


\bibitem[Cheng et~al\mbox{.}(2022)]%
        {cheng2022interest}
\bibfield{author}{\bibinfo{person}{Ruijia Cheng}, \bibinfo{person}{Sayamindu Dasgupta}, {and} \bibinfo{person}{Benjamin~Mako Hill}.} \bibinfo{year}{2022}\natexlab{}.
\newblock \showarticletitle{How Interest-Driven Content Creation Shapes Opportunities for Informal Learning in Scratch: A Case Study on Novices’ Use of Data Structures}. In \bibinfo{booktitle}{\emph{Proceedings of the 2022 CHI Conference on Human Factors in Computing Systems}}. \bibinfo{pages}{1--16}.
\newblock


\bibitem[Cheng and Hill(2022)]%
        {cheng2022many}
\bibfield{author}{\bibinfo{person}{Ruijia Cheng} {and} \bibinfo{person}{Benjamin~Mako Hill}.} \bibinfo{year}{2022}\natexlab{}.
\newblock \showarticletitle{Many destinations, many pathways: A quantitative analysis of legitimate peripheral participation in scratch}.
\newblock \bibinfo{journal}{\emph{Proceedings of the ACM on Human-Computer Interaction}} \bibinfo{volume}{6}, \bibinfo{number}{CSCW2} (\bibinfo{year}{2022}), \bibinfo{pages}{1--26}.
\newblock


\bibitem[Cheng and Zachry(2020)]%
        {cheng2020building}
\bibfield{author}{\bibinfo{person}{Ruijia Cheng} {and} \bibinfo{person}{Mark Zachry}.} \bibinfo{year}{2020}\natexlab{}.
\newblock \showarticletitle{Building community knowledge in online competitions: motivation, practices and challenges}.
\newblock \bibinfo{journal}{\emph{Proceedings of the ACM on Human-Computer Interaction}} \bibinfo{volume}{4}, \bibinfo{number}{CSCW2} (\bibinfo{year}{2020}), \bibinfo{pages}{1--22}.
\newblock


\bibitem[Cheng et~al\mbox{.}(2020)]%
        {cheng2020critique}
\bibfield{author}{\bibinfo{person}{Ruijia Cheng}, \bibinfo{person}{Ziwen Zeng}, \bibinfo{person}{Maysnow Liu}, {and} \bibinfo{person}{Steven Dow}.} \bibinfo{year}{2020}\natexlab{}.
\newblock \showarticletitle{Critique me: exploring how creators publicly request feedback in an online critique community}.
\newblock \bibinfo{journal}{\emph{Proceedings of the ACM on Human-Computer Interaction}} \bibinfo{volume}{4}, \bibinfo{number}{CSCW2} (\bibinfo{year}{2020}), \bibinfo{pages}{1--24}.
\newblock


\bibitem[Cherry and Latulipe(2014)]%
        {cherry2014quantifying}
\bibfield{author}{\bibinfo{person}{Erin Cherry} {and} \bibinfo{person}{Celine Latulipe}.} \bibinfo{year}{2014}\natexlab{}.
\newblock \showarticletitle{Quantifying the creativity support of digital tools through the creativity support index}.
\newblock \bibinfo{journal}{\emph{ACM Transactions on Computer-Human Interaction (TOCHI)}} \bibinfo{volume}{21}, \bibinfo{number}{4} (\bibinfo{year}{2014}), \bibinfo{pages}{1--25}.
\newblock


\bibitem[Chi and Wylie(2014)]%
        {chi2014icap}
\bibfield{author}{\bibinfo{person}{Michelene~TH Chi} {and} \bibinfo{person}{Ruth Wylie}.} \bibinfo{year}{2014}\natexlab{}.
\newblock \showarticletitle{The ICAP framework: Linking cognitive engagement to active learning outcomes}.
\newblock \bibinfo{journal}{\emph{Educational psychologist}} \bibinfo{volume}{49}, \bibinfo{number}{4} (\bibinfo{year}{2014}), \bibinfo{pages}{219--243}.
\newblock


\bibitem[Chiou et~al\mbox{.}(2017)]%
        {chiou2017analyzing}
\bibfield{author}{\bibinfo{person}{Chei-Chang Chiou}, \bibinfo{person}{Li-Tze Lee}, \bibinfo{person}{Li-Chu Tien}, {and} \bibinfo{person}{Yu-Min Wang}.} \bibinfo{year}{2017}\natexlab{}.
\newblock \showarticletitle{Analyzing the effects of various concept mapping techniques on learning achievement under different learning styles}.
\newblock \bibinfo{journal}{\emph{Eurasia Journal of Mathematics, Science and Technology Education}} \bibinfo{volume}{13}, \bibinfo{number}{7} (\bibinfo{year}{2017}), \bibinfo{pages}{3687--3708}.
\newblock


\bibitem[Dasgupta et~al\mbox{.}(2016)]%
        {dasgupta2016remixing}
\bibfield{author}{\bibinfo{person}{Sayamindu Dasgupta}, \bibinfo{person}{William Hale}, \bibinfo{person}{Andr{\'e}s Monroy-Hern{\'a}ndez}, {and} \bibinfo{person}{Benjamin~Mako Hill}.} \bibinfo{year}{2016}\natexlab{}.
\newblock \showarticletitle{Remixing as a pathway to computational thinking}. In \bibinfo{booktitle}{\emph{Proceedings of the 19th ACM Conference on Computer-Supported Cooperative Work \& Social Computing}}. \bibinfo{pages}{1438--1449}.
\newblock


\bibitem[Dasgupta and Hill(2017)]%
        {dasgupta2017scratch}
\bibfield{author}{\bibinfo{person}{Sayamindu Dasgupta} {and} \bibinfo{person}{Benjamin~Mako Hill}.} \bibinfo{year}{2017}\natexlab{}.
\newblock \showarticletitle{Scratch community blocks: Supporting children as data scientists}. In \bibinfo{booktitle}{\emph{Proceedings of the 2017 CHI conference on human factors in computing systems}}. \bibinfo{pages}{3620--3631}.
\newblock


\bibitem[Deiner et~al\mbox{.}(2023)]%
        {deiner2023automated}
\bibfield{author}{\bibinfo{person}{Adina Deiner}, \bibinfo{person}{Patric Feldmeier}, \bibinfo{person}{Gordon Fraser}, \bibinfo{person}{Sebastian Schweikl}, {and} \bibinfo{person}{Wengran Wang}.} \bibinfo{year}{2023}\natexlab{}.
\newblock \showarticletitle{Automated test generation for Scratch programs}.
\newblock \bibinfo{journal}{\emph{Empirical Software Engineering}} \bibinfo{volume}{28}, \bibinfo{number}{3} (\bibinfo{year}{2023}), \bibinfo{pages}{79}.
\newblock


\bibitem[Dietz et~al\mbox{.}(2021)]%
        {dietz2021storycoder}
\bibfield{author}{\bibinfo{person}{Griffin Dietz}, \bibinfo{person}{Jimmy~K Le}, \bibinfo{person}{Nadin Tamer}, \bibinfo{person}{Jenny Han}, \bibinfo{person}{Hyowon Gweon}, \bibinfo{person}{Elizabeth~L Murnane}, {and} \bibinfo{person}{James~A Landay}.} \bibinfo{year}{2021}\natexlab{}.
\newblock \showarticletitle{Storycoder: Teaching computational thinking concepts through storytelling in a voice-guided app for children}. In \bibinfo{booktitle}{\emph{Proceedings of the 2021 CHI Conference on Human Factors in Computing Systems}}. \bibinfo{pages}{1--15}.
\newblock


\bibitem[Dietz et~al\mbox{.}(2023)]%
        {dietz2023visual}
\bibfield{author}{\bibinfo{person}{Griffin Dietz}, \bibinfo{person}{Nadin Tamer}, \bibinfo{person}{Carina Ly}, \bibinfo{person}{Jimmy~K Le}, {and} \bibinfo{person}{James~A Landay}.} \bibinfo{year}{2023}\natexlab{}.
\newblock \showarticletitle{Visual StoryCoder: A Multimodal Programming Environment for Children’s Creation of Stories}. In \bibinfo{booktitle}{\emph{Proceedings of the 2023 CHI Conference on Human Factors in Computing Systems}}. \bibinfo{pages}{1--16}.
\newblock


\bibitem[Fagerlund et~al\mbox{.}(2021)]%
        {fagerlund2021computational}
\bibfield{author}{\bibinfo{person}{Janne Fagerlund}, \bibinfo{person}{P{\"a}ivi H{\"a}kkinen}, \bibinfo{person}{Mikko Vesisenaho}, {and} \bibinfo{person}{Jouni Viiri}.} \bibinfo{year}{2021}\natexlab{}.
\newblock \showarticletitle{Computational thinking in programming with Scratch in primary schools: A systematic review}.
\newblock \bibinfo{journal}{\emph{Computer Applications in Engineering Education}} \bibinfo{volume}{29}, \bibinfo{number}{1} (\bibinfo{year}{2021}), \bibinfo{pages}{12--28}.
\newblock


\bibitem[Fan et~al\mbox{.}(2024)]%
        {fan2024survey}
\bibfield{author}{\bibinfo{person}{Wenqi Fan}, \bibinfo{person}{Yujuan Ding}, \bibinfo{person}{Liangbo Ning}, \bibinfo{person}{Shijie Wang}, \bibinfo{person}{Hengyun Li}, \bibinfo{person}{Dawei Yin}, \bibinfo{person}{Tat-Seng Chua}, {and} \bibinfo{person}{Qing Li}.} \bibinfo{year}{2024}\natexlab{}.
\newblock \showarticletitle{A Survey on RAG Meeting LLMs: Towards Retrieval-Augmented Large Language Models}. In \bibinfo{booktitle}{\emph{Proceedings of the 30th ACM SIGKDD Conference on Knowledge Discovery and Data Mining}}. \bibinfo{pages}{6491--6501}.
\newblock


\bibitem[Fields et~al\mbox{.}(2014)]%
        {fields2014programming}
\bibfield{author}{\bibinfo{person}{Deborah~A Fields}, \bibinfo{person}{Michael Giang}, {and} \bibinfo{person}{Yasmin Kafai}.} \bibinfo{year}{2014}\natexlab{}.
\newblock \showarticletitle{Programming in the wild: trends in youth computational participation in the online scratch community}. In \bibinfo{booktitle}{\emph{Proceedings of the 9th workshop in primary and secondary computing education}}. \bibinfo{pages}{2--11}.
\newblock


\bibitem[Foong et~al\mbox{.}(2017)]%
        {foong2017online}
\bibfield{author}{\bibinfo{person}{Eureka Foong}, \bibinfo{person}{Steven~P Dow}, \bibinfo{person}{Brian~P Bailey}, {and} \bibinfo{person}{Elizabeth~M Gerber}.} \bibinfo{year}{2017}\natexlab{}.
\newblock \showarticletitle{Online feedback exchange: A framework for understanding the socio-psychological factors}. In \bibinfo{booktitle}{\emph{Proceedings of the 2017 CHI Conference on Human Factors in Computing Systems}}. \bibinfo{pages}{4454--4467}.
\newblock


\bibitem[Ford et~al\mbox{.}(2018)]%
        {ford2018we}
\bibfield{author}{\bibinfo{person}{Denae Ford}, \bibinfo{person}{Kristina Lustig}, \bibinfo{person}{Jeremy Banks}, {and} \bibinfo{person}{Chris Parnin}.} \bibinfo{year}{2018}\natexlab{}.
\newblock \showarticletitle{" We Don't Do That Here" How Collaborative Editing with Mentors Improves Engagement in Social Q\&A Communities}. In \bibinfo{booktitle}{\emph{Proceedings of the 2018 CHI conference on human factors in computing systems}}. \bibinfo{pages}{1--12}.
\newblock


\bibitem[Ford et~al\mbox{.}(2016)]%
        {ford2016paradise}
\bibfield{author}{\bibinfo{person}{Denae Ford}, \bibinfo{person}{Justin Smith}, \bibinfo{person}{Philip~J Guo}, {and} \bibinfo{person}{Chris Parnin}.} \bibinfo{year}{2016}\natexlab{}.
\newblock \showarticletitle{Paradise unplugged: Identifying barriers for female participation on stack overflow}. In \bibinfo{booktitle}{\emph{Proceedings of the 2016 24th ACM SIGSOFT International symposium on foundations of software engineering}}. \bibinfo{pages}{846--857}.
\newblock


\bibitem[Funke and Geldreich(2017)]%
        {funke2017gender}
\bibfield{author}{\bibinfo{person}{Alexandra Funke} {and} \bibinfo{person}{Katharina Geldreich}.} \bibinfo{year}{2017}\natexlab{}.
\newblock \showarticletitle{Gender differences in scratch programs of primary school children}. In \bibinfo{booktitle}{\emph{Proceedings of the 12th workshop on primary and secondary computing education}}. \bibinfo{pages}{57--64}.
\newblock


\bibitem[Gabajiwala et~al\mbox{.}(2022)]%
        {gabajiwala2022quiz}
\bibfield{author}{\bibinfo{person}{Ebrahim Gabajiwala}, \bibinfo{person}{Priyav Mehta}, \bibinfo{person}{Ritik Singh}, {and} \bibinfo{person}{Reeta Koshy}.} \bibinfo{year}{2022}\natexlab{}.
\newblock \showarticletitle{Quiz maker: Automatic quiz generation from text using NLP}. In \bibinfo{booktitle}{\emph{Futuristic Trends in Networks and Computing Technologies: Select Proceedings of Fourth International Conference on FTNCT 2021}}. Springer, \bibinfo{pages}{523--533}.
\newblock


\bibitem[Gan et~al\mbox{.}(2018)]%
        {gan2018gender}
\bibfield{author}{\bibinfo{person}{Emilia~F Gan}, \bibinfo{person}{Benjamin~Mako Hill}, {and} \bibinfo{person}{Sayamindu Dasgupta}.} \bibinfo{year}{2018}\natexlab{}.
\newblock \showarticletitle{Gender, Feedback, and Learners' Decisions to Share Their Creative Computing Projects}.
\newblock \bibinfo{journal}{\emph{Proceedings of the ACM on Human-Computer Interaction}} \bibinfo{volume}{2}, \bibinfo{number}{CSCW} (\bibinfo{year}{2018}), \bibinfo{pages}{1--23}.
\newblock


\bibitem[G{\"u}nbatar(2019)]%
        {gunbatar2019computational}
\bibfield{author}{\bibinfo{person}{Mustafa~Serkan G{\"u}nbatar}.} \bibinfo{year}{2019}\natexlab{}.
\newblock \showarticletitle{Computational thinking within the context of professional life: Change in CT skill from the viewpoint of teachers}.
\newblock \bibinfo{journal}{\emph{Education and Information Technologies}} \bibinfo{volume}{24}, \bibinfo{number}{5} (\bibinfo{year}{2019}), \bibinfo{pages}{2629--2652}.
\newblock


\bibitem[Guo et~al\mbox{.}(2023)]%
        {guo2023makes}
\bibfield{author}{\bibinfo{person}{Qingyu Guo}, \bibinfo{person}{Chao Zhang}, \bibinfo{person}{Hanfang Lyu}, \bibinfo{person}{Zhenhui Peng}, {and} \bibinfo{person}{Xiaojuan Ma}.} \bibinfo{year}{2023}\natexlab{}.
\newblock \showarticletitle{What Makes Creators Engage with Online Critiques? Understanding the Role of Artifacts’ Creation Stage, Characteristics of Community Comments, and their Interactions}. In \bibinfo{booktitle}{\emph{Proceedings of the 2023 CHI Conference on Human Factors in Computing Systems}}. \bibinfo{pages}{1--17}.
\newblock


\bibitem[Hill and Monroy-Hern{\'a}ndez(2013)]%
        {hill2013remixing}
\bibfield{author}{\bibinfo{person}{Benjamin~Mako Hill} {and} \bibinfo{person}{Andr{\'e}s Monroy-Hern{\'a}ndez}.} \bibinfo{year}{2013}\natexlab{}.
\newblock \showarticletitle{The remixing dilemma: The trade-off between generativity and originality}.
\newblock \bibinfo{journal}{\emph{American Behavioral Scientist}} \bibinfo{volume}{57}, \bibinfo{number}{5} (\bibinfo{year}{2013}), \bibinfo{pages}{643--663}.
\newblock


\bibitem[Hsu(2014)]%
        {hsu2014gender}
\bibfield{author}{\bibinfo{person}{Hui-mei~Justina Hsu}.} \bibinfo{year}{2014}\natexlab{}.
\newblock \showarticletitle{Gender differences in Scratch Game design}. In \bibinfo{booktitle}{\emph{2014 International Conference on Information, Business and Education Technology (ICIBET 2014)}}. Atlantis Press, \bibinfo{pages}{100--103}.
\newblock


\bibitem[Ji et~al\mbox{.}(2023)]%
        {ji2023systematic}
\bibfield{author}{\bibinfo{person}{Hyangeun Ji}, \bibinfo{person}{Insook Han}, {and} \bibinfo{person}{Yujung Ko}.} \bibinfo{year}{2023}\natexlab{}.
\newblock \showarticletitle{A systematic review of conversational AI in language education: Focusing on the collaboration with human teachers}.
\newblock \bibinfo{journal}{\emph{Journal of Research on Technology in Education}} \bibinfo{volume}{55}, \bibinfo{number}{1} (\bibinfo{year}{2023}), \bibinfo{pages}{48--63}.
\newblock


\bibitem[Jia et~al\mbox{.}(2021)]%
        {jia2021all}
\bibfield{author}{\bibinfo{person}{Qinjin Jia}, \bibinfo{person}{Jialin Cui}, \bibinfo{person}{Yunkai Xiao}, \bibinfo{person}{Chengyuan Liu}, \bibinfo{person}{Parvez Rashid}, {and} \bibinfo{person}{Edward~F Gehringer}.} \bibinfo{year}{2021}\natexlab{}.
\newblock \showarticletitle{All-in-one: Multi-task learning bert models for evaluating peer assessments}.
\newblock \bibinfo{journal}{\emph{arXiv preprint arXiv:2110.03895}} (\bibinfo{year}{2021}).
\newblock


\bibitem[Jin et~al\mbox{.}(2024)]%
        {jin2024teach}
\bibfield{author}{\bibinfo{person}{Hyoungwook Jin}, \bibinfo{person}{Seonghee Lee}, \bibinfo{person}{Hyungyu Shin}, {and} \bibinfo{person}{Juho Kim}.} \bibinfo{year}{2024}\natexlab{}.
\newblock \showarticletitle{Teach AI How to Code: Using Large Language Models as Teachable Agents for Programming Education}. In \bibinfo{booktitle}{\emph{Proceedings of the CHI Conference on Human Factors in Computing Systems}}. \bibinfo{pages}{1--28}.
\newblock


\bibitem[Kapoor(2023)]%
        {Kapoor2023NegativePrompts}
\bibfield{author}{\bibinfo{person}{Mukund Kapoor}.} \bibinfo{year}{2023}\natexlab{}.
\newblock \bibinfo{title}{Negative Prompts in Stable Diffusion: A Beginner's Guide}.
\newblock
\newblock
\urldef\tempurl%
\url{https://www.greataiprompts.com/imageprompt/what-is-negative-prompt-in-stable-diffusion/}
\showURL{%
\tempurl}
\newblock
\shownote{Accessed: 2023-12-16}.


\bibitem[Kasneci et~al\mbox{.}(2023)]%
        {kasneci2023chatgpt}
\bibfield{author}{\bibinfo{person}{Enkelejda Kasneci}, \bibinfo{person}{Kathrin Se{\ss}ler}, \bibinfo{person}{Stefan K{\"u}chemann}, \bibinfo{person}{Maria Bannert}, \bibinfo{person}{Daryna Dementieva}, \bibinfo{person}{Frank Fischer}, \bibinfo{person}{Urs Gasser}, \bibinfo{person}{Georg Groh}, \bibinfo{person}{Stephan G{\"u}nnemann}, \bibinfo{person}{Eyke H{\"u}llermeier}, {et~al\mbox{.}}} \bibinfo{year}{2023}\natexlab{}.
\newblock \showarticletitle{ChatGPT for good? On opportunities and challenges of large language models for education}.
\newblock \bibinfo{journal}{\emph{Learning and individual differences}}  \bibinfo{volume}{103} (\bibinfo{year}{2023}), \bibinfo{pages}{102274}.
\newblock


\bibitem[Kazemitabaar et~al\mbox{.}(2024)]%
        {kazemitabaar2024codeaid}
\bibfield{author}{\bibinfo{person}{Majeed Kazemitabaar}, \bibinfo{person}{Runlong Ye}, \bibinfo{person}{Xiaoning Wang}, \bibinfo{person}{Austin~Zachary Henley}, \bibinfo{person}{Paul Denny}, \bibinfo{person}{Michelle Craig}, {and} \bibinfo{person}{Tovi Grossman}.} \bibinfo{year}{2024}\natexlab{}.
\newblock \showarticletitle{Codeaid: Evaluating a classroom deployment of an llm-based programming assistant that balances student and educator needs}. In \bibinfo{booktitle}{\emph{Proceedings of the CHI Conference on Human Factors in Computing Systems}}. \bibinfo{pages}{1--20}.
\newblock


\bibitem[Knezic et~al\mbox{.}(2010)]%
        {knezic2010socratic}
\bibfield{author}{\bibinfo{person}{Dubravka Knezic}, \bibinfo{person}{Theo Wubbels}, \bibinfo{person}{Ed Elbers}, {and} \bibinfo{person}{Maaike Hajer}.} \bibinfo{year}{2010}\natexlab{}.
\newblock \showarticletitle{The Socratic Dialogue and teacher education}.
\newblock \bibinfo{journal}{\emph{Teaching and teacher education}} \bibinfo{volume}{26}, \bibinfo{number}{4} (\bibinfo{year}{2010}), \bibinfo{pages}{1104--1111}.
\newblock


\bibitem[Korkmaz et~al\mbox{.}(2017)]%
        {korkmaz2017validity}
\bibfield{author}{\bibinfo{person}{{\"O}zgen Korkmaz}, \bibinfo{person}{Recep {\c{C}}akir}, {and} \bibinfo{person}{M~Ya{\c{s}}ar {\"O}zden}.} \bibinfo{year}{2017}\natexlab{}.
\newblock \showarticletitle{A validity and reliability study of the computational thinking scales (CTS)}.
\newblock \bibinfo{journal}{\emph{Computers in human behavior}}  \bibinfo{volume}{72} (\bibinfo{year}{2017}), \bibinfo{pages}{558--569}.
\newblock


\bibitem[Ladias et~al\mbox{.}(2021)]%
        {ladias2021codeorama}
\bibfield{author}{\bibinfo{person}{Anastasios Ladias}, \bibinfo{person}{Aristotelis Mikropoulos}, \bibinfo{person}{Demetrios Ladias}, {and} \bibinfo{person}{Ioanna Bellou}.} \bibinfo{year}{2021}\natexlab{}.
\newblock \showarticletitle{CodeOrama: A Two-Dimensional Visualization Tool for Scratch Code to Assist Young Learners' Understanding of Computer Programming.}
\newblock \bibinfo{journal}{\emph{Themes in eLearning}}  \bibinfo{volume}{14} (\bibinfo{year}{2021}), \bibinfo{pages}{31--41}.
\newblock


\bibitem[Lave and Wenger(1991)]%
        {lave1991situated}
\bibfield{author}{\bibinfo{person}{Jean Lave} {and} \bibinfo{person}{Etienne Wenger}.} \bibinfo{year}{1991}\natexlab{}.
\newblock \bibinfo{booktitle}{\emph{Situated learning: Legitimate peripheral participation}}.
\newblock \bibinfo{publisher}{Cambridge university press}.
\newblock


\bibitem[Lin et~al\mbox{.}(2022)]%
        {lin2022development}
\bibfield{author}{\bibinfo{person}{Peng-Chun Lin}, \bibinfo{person}{Huei-Tse Hou}, {and} \bibinfo{person}{Kuo-En Chang}.} \bibinfo{year}{2022}\natexlab{}.
\newblock \showarticletitle{The development of a collaborative problem solving environment that integrates a scaffolding mind tool and simulation-based learning: An analysis of learners’ performance and their cognitive process in discussion}.
\newblock \bibinfo{journal}{\emph{Interactive Learning Environments}} \bibinfo{volume}{30}, \bibinfo{number}{7} (\bibinfo{year}{2022}), \bibinfo{pages}{1273--1290}.
\newblock


\bibitem[Liu et~al\mbox{.}(2024)]%
        {liu2024exploring}
\bibfield{author}{\bibinfo{person}{Fang Liu}, \bibinfo{person}{Yang Liu}, \bibinfo{person}{Lin Shi}, \bibinfo{person}{Houkun Huang}, \bibinfo{person}{Ruifeng Wang}, \bibinfo{person}{Zhen Yang}, {and} \bibinfo{person}{Li Zhang}.} \bibinfo{year}{2024}\natexlab{}.
\newblock \showarticletitle{Exploring and evaluating hallucinations in llm-powered code generation}.
\newblock \bibinfo{journal}{\emph{arXiv preprint arXiv:2404.00971}} (\bibinfo{year}{2024}).
\newblock


\bibitem[Lovell et~al\mbox{.}(2023)]%
        {lovell2023lilytiny}
\bibfield{author}{\bibinfo{person}{Emily Lovell}, \bibinfo{person}{Leah Buechley}, {and} \bibinfo{person}{James Davis}.} \bibinfo{year}{2023}\natexlab{}.
\newblock \showarticletitle{LilyTiny in the Wild: Studying the Adoption of a Low-Cost Sewable Microcontroller for Computing Education}. In \bibinfo{booktitle}{\emph{Proceedings of the 2023 ACM Designing Interactive Systems Conference}}. \bibinfo{pages}{282--293}.
\newblock


\bibitem[Malycha and Maier(2017)]%
        {malycha2017enhancing}
\bibfield{author}{\bibinfo{person}{Charlotte~P Malycha} {and} \bibinfo{person}{G{\"u}nter~W Maier}.} \bibinfo{year}{2017}\natexlab{}.
\newblock \showarticletitle{Enhancing creativity on different complexity levels by eliciting mental models.}
\newblock \bibinfo{journal}{\emph{Psychology of Aesthetics, Creativity, and the Arts}} \bibinfo{volume}{11}, \bibinfo{number}{2} (\bibinfo{year}{2017}), \bibinfo{pages}{187}.
\newblock


\bibitem[Matias et~al\mbox{.}(2016)]%
        {matias2016skill}
\bibfield{author}{\bibinfo{person}{J~Nathan Matias}, \bibinfo{person}{Sayamindu Dasgupta}, {and} \bibinfo{person}{Benjamin~Mako Hill}.} \bibinfo{year}{2016}\natexlab{}.
\newblock \showarticletitle{Skill progression in scratch revisited}. In \bibinfo{booktitle}{\emph{Proceedings of the 2016 CHI conference on human factors in computing systems}}. \bibinfo{pages}{1486--1490}.
\newblock


\bibitem[Monroy-Hern{\'a}ndez(2007)]%
        {monroy2007scratchr}
\bibfield{author}{\bibinfo{person}{Andr{\'e}s Monroy-Hern{\'a}ndez}.} \bibinfo{year}{2007}\natexlab{}.
\newblock \showarticletitle{ScratchR: sharing user-generated programmable media}. In \bibinfo{booktitle}{\emph{Proceedings of the 6th international conference on Interaction design and children}}. \bibinfo{pages}{167--168}.
\newblock


\bibitem[Moreno-Le{\'o}n and Robles(2015)]%
        {moreno2015dr}
\bibfield{author}{\bibinfo{person}{Jes{\'u}s Moreno-Le{\'o}n} {and} \bibinfo{person}{Gregorio Robles}.} \bibinfo{year}{2015}\natexlab{}.
\newblock \showarticletitle{Dr. Scratch: A web tool to automatically evaluate Scratch projects}. In \bibinfo{booktitle}{\emph{Proceedings of the workshop in primary and secondary computing education}}. \bibinfo{publisher}{Association for Computing Machinery}, \bibinfo{address}{New York, NY, USA}, \bibinfo{pages}{132--133}.
\newblock


\bibitem[Morrison et~al\mbox{.}(2014)]%
        {morrison2014measuring}
\bibfield{author}{\bibinfo{person}{Briana~B Morrison}, \bibinfo{person}{Brian Dorn}, {and} \bibinfo{person}{Mark Guzdial}.} \bibinfo{year}{2014}\natexlab{}.
\newblock \showarticletitle{Measuring cognitive load in introductory CS: adaptation of an instrument}. In \bibinfo{booktitle}{\emph{Proceedings of the tenth annual conference on International computing education research}}. \bibinfo{pages}{131--138}.
\newblock


\bibitem[Online(2023)]%
        {StableDiffusion2023}
\bibfield{author}{\bibinfo{person}{Stable~Diffusion Online}.} \bibinfo{year}{2023}\natexlab{}.
\newblock \bibinfo{title}{Stable Diffusion: A Latent Text-to-Image Diffusion Model}.
\newblock
\newblock
\urldef\tempurl%
\url{https://stablediffusionweb.com/}
\showURL{%
\tempurl}
\newblock
\shownote{Accessed: 2023-09-08}.


\bibitem[Oppenlaender et~al\mbox{.}(2020)]%
        {oppenlaender2020crowdui}
\bibfield{author}{\bibinfo{person}{Jonas Oppenlaender}, \bibinfo{person}{Thanassis Tiropanis}, {and} \bibinfo{person}{Simo Hosio}.} \bibinfo{year}{2020}\natexlab{}.
\newblock \showarticletitle{CrowdUI: Supporting web design with the crowd}.
\newblock \bibinfo{journal}{\emph{Proceedings of the ACM on Human-Computer Interaction}} \bibinfo{volume}{4}, \bibinfo{number}{EICS} (\bibinfo{year}{2020}), \bibinfo{pages}{1--28}.
\newblock


\bibitem[Park et~al\mbox{.}(2017)]%
        {park2017telling}
\bibfield{author}{\bibinfo{person}{Hae~Won Park}, \bibinfo{person}{Mirko Gelsomini}, \bibinfo{person}{Jin~Joo Lee}, {and} \bibinfo{person}{Cynthia Breazeal}.} \bibinfo{year}{2017}\natexlab{}.
\newblock \showarticletitle{Telling stories to robots: The effect of backchanneling on a child's storytelling}. In \bibinfo{booktitle}{\emph{Proceedings of the 2017 ACM/IEEE international conference on human-robot interaction}}. \bibinfo{pages}{100--108}.
\newblock


\bibitem[Peng et~al\mbox{.}(2024)]%
        {peng2024designquizzer}
\bibfield{author}{\bibinfo{person}{Zhenhui Peng}, \bibinfo{person}{Qiaoyi Chen}, \bibinfo{person}{Zhiyu Shen}, \bibinfo{person}{Xiaojuan Ma}, {and} \bibinfo{person}{Antti Oulasvirta}.} \bibinfo{year}{2024}\natexlab{}.
\newblock \showarticletitle{DesignQuizzer: A Community-Powered Conversational Agent for Learning Visual Design}.
\newblock \bibinfo{journal}{\emph{Proceedings of the ACM on Human-Computer Interaction}} \bibinfo{volume}{8}, \bibinfo{number}{CSCW1} (\bibinfo{year}{2024}), \bibinfo{pages}{1--40}.
\newblock


\bibitem[Peng et~al\mbox{.}(2020)]%
        {peng2020exploring}
\bibfield{author}{\bibinfo{person}{Zhenhui Peng}, \bibinfo{person}{Qingyu Guo}, \bibinfo{person}{Ka~Wing Tsang}, {and} \bibinfo{person}{Xiaojuan Ma}.} \bibinfo{year}{2020}\natexlab{}.
\newblock \showarticletitle{Exploring the effects of technological writing assistance for support providers in online mental health community}. In \bibinfo{booktitle}{\emph{Proceedings of the 2020 CHI Conference on Human Factors in Computing Systems}}. \bibinfo{pages}{1--15}.
\newblock


\bibitem[Resch et~al\mbox{.}(2024)]%
        {resch2024overcoming}
\bibfield{author}{\bibinfo{person}{Katharina Resch}, \bibinfo{person}{Ilse Schrittesser}, {and} \bibinfo{person}{Mariella Knapp}.} \bibinfo{year}{2024}\natexlab{}.
\newblock \showarticletitle{Overcoming the theory-practice divide in teacher education with the ‘Partner School Programme’. A conceptual mapping}.
\newblock \bibinfo{journal}{\emph{European Journal of Teacher Education}} \bibinfo{volume}{47}, \bibinfo{number}{3} (\bibinfo{year}{2024}), \bibinfo{pages}{564--580}.
\newblock


\bibitem[Resnick et~al\mbox{.}(2009)]%
        {10.1145/1592761.1592779}
\bibfield{author}{\bibinfo{person}{Mitchel Resnick}, \bibinfo{person}{John Maloney}, \bibinfo{person}{Andr\'{e}s Monroy-Hern\'{a}ndez}, \bibinfo{person}{Natalie Rusk}, \bibinfo{person}{Evelyn Eastmond}, \bibinfo{person}{Karen Brennan}, \bibinfo{person}{Amon Millner}, \bibinfo{person}{Eric Rosenbaum}, \bibinfo{person}{Jay Silver}, \bibinfo{person}{Brian Silverman}, {and} \bibinfo{person}{Yasmin Kafai}.} \bibinfo{year}{2009}\natexlab{}.
\newblock \showarticletitle{Scratch: Programming for All}.
\newblock \bibinfo{journal}{\emph{Commun. ACM}} \bibinfo{volume}{52}, \bibinfo{number}{11} (\bibinfo{date}{nov} \bibinfo{year}{2009}), \bibinfo{pages}{60–67}.
\newblock
\showISSN{0001-0782}
\urldef\tempurl%
\url{https://doi.org/10.1145/1592761.1592779}
\showDOI{\tempurl}


\bibitem[Richard and Kafai(2016)]%
        {richard2016blind}
\bibfield{author}{\bibinfo{person}{Gabriela~T Richard} {and} \bibinfo{person}{Yasmin~B Kafai}.} \bibinfo{year}{2016}\natexlab{}.
\newblock \showarticletitle{Blind spots in youth DIY programming: Examining diversity in creators, content, and comments within the scratch online community}. In \bibinfo{booktitle}{\emph{Proceedings of the 2016 CHI conference on Human Factors in Computing Systems}}. \bibinfo{pages}{1473--1485}.
\newblock


\bibitem[Romero and Artal-Sevil(2021)]%
        {romero2021scratch}
\bibfield{author}{\bibinfo{person}{E Romero} {and} \bibinfo{person}{JS Artal-Sevil}.} \bibinfo{year}{2021}\natexlab{}.
\newblock \showarticletitle{Scratch: a tool for learning and teaching}. In \bibinfo{booktitle}{\emph{ICERI2021 Proceedings}}. IATED, \bibinfo{pages}{7844--7853}.
\newblock


\bibitem[S{\'a}ez-L{\'o}pez et~al\mbox{.}(2016)]%
        {saez2016visual}
\bibfield{author}{\bibinfo{person}{Jos{\'e}-Manuel S{\'a}ez-L{\'o}pez}, \bibinfo{person}{Marcos Rom{\'a}n-Gonz{\'a}lez}, {and} \bibinfo{person}{Esteban V{\'a}zquez-Cano}.} \bibinfo{year}{2016}\natexlab{}.
\newblock \showarticletitle{Visual programming languages integrated across the curriculum in elementary school: A two year case study using “Scratch” in five schools}.
\newblock \bibinfo{journal}{\emph{Computers \& Education}}  \bibinfo{volume}{97} (\bibinfo{year}{2016}), \bibinfo{pages}{129--141}.
\newblock


\bibitem[Seo et~al\mbox{.}(2024)]%
        {seo2024chacha}
\bibfield{author}{\bibinfo{person}{Woosuk Seo}, \bibinfo{person}{Chanmo Yang}, {and} \bibinfo{person}{Young-Ho Kim}.} \bibinfo{year}{2024}\natexlab{}.
\newblock \showarticletitle{ChaCha: Leveraging Large Language Models to Prompt Children to Share Their Emotions about Personal Events}. In \bibinfo{booktitle}{\emph{Proceedings of the CHI Conference on Human Factors in Computing Systems}}. \bibinfo{pages}{1--20}.
\newblock


\bibitem[Shahriar and Matsuda(2021)]%
        {shahriar2021can}
\bibfield{author}{\bibinfo{person}{Tasmia Shahriar} {and} \bibinfo{person}{Noboru Matsuda}.} \bibinfo{year}{2021}\natexlab{}.
\newblock \showarticletitle{“Can you clarify what you said?”: Studying the impact of tutee agents’ follow-up questions on tutors’ learning}. In \bibinfo{booktitle}{\emph{Artificial Intelligence in Education: 22nd International Conference, AIED 2021, Utrecht, The Netherlands, June 14--18, 2021, Proceedings, Part I 22}}. Springer, \bibinfo{pages}{395--407}.
\newblock


\bibitem[Shanahan et~al\mbox{.}(2023)]%
        {Shanahan2023Role}
\bibfield{author}{\bibinfo{person}{Murray Shanahan}, \bibinfo{person}{Kyle McDonell}, {and} \bibinfo{person}{Laria Reynolds}.} \bibinfo{year}{2023}\natexlab{}.
\newblock \showarticletitle{Role play with large language models}.
\newblock \bibinfo{journal}{\emph{Nature}}  \bibinfo{volume}{623} (\bibinfo{year}{2023}), \bibinfo{pages}{493--498}.
\newblock
\urldef\tempurl%
\url{https://doi.org/10.1038/s41586-023-06647-8}
\showDOI{\tempurl}


\bibitem[Shorey et~al\mbox{.}(2021)]%
        {shorey2021hanging}
\bibfield{author}{\bibinfo{person}{Samantha Shorey}, \bibinfo{person}{Benjamin~Mako Hill}, {and} \bibinfo{person}{Samuel Woolley}.} \bibinfo{year}{2021}\natexlab{}.
\newblock \showarticletitle{From hanging out to figuring it out: Socializing online as a pathway to computational thinking}.
\newblock \bibinfo{journal}{\emph{New Media \& Society}} \bibinfo{volume}{23}, \bibinfo{number}{8} (\bibinfo{year}{2021}), \bibinfo{pages}{2327--2344}.
\newblock


\bibitem[Stone(2023)]%
        {Stone2023NegativeAIPrompts}
\bibfield{author}{\bibinfo{person}{Don Stone}.} \bibinfo{year}{2023}\natexlab{}.
\newblock \bibinfo{title}{What Do Negative AI Prompts Do and How They’ll Improve Your AI Art}.
\newblock
\newblock
\urldef\tempurl%
\url{https://interlinkedai.com/what-do-negative-ai-prompts-do-and-how-theyll-improve-your-ai-art/}
\showURL{%
\tempurl}
\newblock
\shownote{Accessed: 2023-12-16}.


\bibitem[Sun et~al\mbox{.}(2022)]%
        {sun2022students}
\bibfield{author}{\bibinfo{person}{Meng Sun}, \bibinfo{person}{Minhong Wang}, \bibinfo{person}{Rupert Wegerif}, {and} \bibinfo{person}{Jun Peng}.} \bibinfo{year}{2022}\natexlab{}.
\newblock \showarticletitle{How do students generate ideas together in scientific creativity tasks through computer-based mind mapping?}
\newblock \bibinfo{journal}{\emph{Computers \& Education}}  \bibinfo{volume}{176} (\bibinfo{year}{2022}), \bibinfo{pages}{104359}.
\newblock


\bibitem[Sweller(1988)]%
        {sweller1988cognitive}
\bibfield{author}{\bibinfo{person}{John Sweller}.} \bibinfo{year}{1988}\natexlab{}.
\newblock \showarticletitle{Cognitive load during problem solving: Effects on learning}.
\newblock \bibinfo{journal}{\emph{Cognitive science}} \bibinfo{volume}{12}, \bibinfo{number}{2} (\bibinfo{year}{1988}), \bibinfo{pages}{257--285}.
\newblock


\bibitem[Tai and Chen(2023)]%
        {tai2023impact}
\bibfield{author}{\bibinfo{person}{Tzu-Yu Tai} {and} \bibinfo{person}{Howard Hao-Jan Chen}.} \bibinfo{year}{2023}\natexlab{}.
\newblock \showarticletitle{The impact of Google Assistant on adolescent EFL learners’ willingness to communicate}.
\newblock \bibinfo{journal}{\emph{Interactive Learning Environments}} \bibinfo{volume}{31}, \bibinfo{number}{3} (\bibinfo{year}{2023}), \bibinfo{pages}{1485--1502}.
\newblock


\bibitem[Tang et~al\mbox{.}(2020)]%
        {tang2020assessing}
\bibfield{author}{\bibinfo{person}{Xiaodan Tang}, \bibinfo{person}{Yue Yin}, \bibinfo{person}{Qiao Lin}, \bibinfo{person}{Roxana Hadad}, {and} \bibinfo{person}{Xiaoming Zhai}.} \bibinfo{year}{2020}\natexlab{}.
\newblock \showarticletitle{Assessing computational thinking: A systematic review of empirical studies}.
\newblock \bibinfo{journal}{\emph{Computers \& Education}}  \bibinfo{volume}{148} (\bibinfo{year}{2020}), \bibinfo{pages}{103798}.
\newblock


\bibitem[Tausczik and Wang(2017)]%
        {tausczik2017share}
\bibfield{author}{\bibinfo{person}{Yla Tausczik} {and} \bibinfo{person}{Ping Wang}.} \bibinfo{year}{2017}\natexlab{}.
\newblock \showarticletitle{To share, or not to share? Community-level collaboration in open innovation contests}.
\newblock \bibinfo{journal}{\emph{Proceedings of the ACM on Human-Computer Interaction}} \bibinfo{volume}{1}, \bibinfo{number}{CSCW} (\bibinfo{year}{2017}), \bibinfo{pages}{1--23}.
\newblock


\bibitem[Tausczik et~al\mbox{.}(2014)]%
        {tausczik2014collaborative}
\bibfield{author}{\bibinfo{person}{Yla~R Tausczik}, \bibinfo{person}{Aniket Kittur}, {and} \bibinfo{person}{Robert~E Kraut}.} \bibinfo{year}{2014}\natexlab{}.
\newblock \showarticletitle{Collaborative problem solving: A study of mathoverflow}. In \bibinfo{booktitle}{\emph{Proceedings of the 17th ACM conference on Computer supported cooperative work \& social computing}}. \bibinfo{pages}{355--367}.
\newblock


\bibitem[Wang et~al\mbox{.}(2021)]%
        {wang2021cass}
\bibfield{author}{\bibinfo{person}{Liuping Wang}, \bibinfo{person}{Dakuo Wang}, \bibinfo{person}{Feng Tian}, \bibinfo{person}{Zhenhui Peng}, \bibinfo{person}{Xiangmin Fan}, \bibinfo{person}{Zhan Zhang}, \bibinfo{person}{Mo Yu}, \bibinfo{person}{Xiaojuan Ma}, {and} \bibinfo{person}{Hongan Wang}.} \bibinfo{year}{2021}\natexlab{}.
\newblock \showarticletitle{Cass: Towards building a social-support chatbot for online health community}.
\newblock \bibinfo{journal}{\emph{Proceedings of the ACM on Human-Computer Interaction}} \bibinfo{volume}{5}, \bibinfo{number}{CSCW1} (\bibinfo{year}{2021}), \bibinfo{pages}{1--31}.
\newblock


\bibitem[Winkler et~al\mbox{.}(2020)]%
        {winkler2020sara}
\bibfield{author}{\bibinfo{person}{Rainer Winkler}, \bibinfo{person}{Sebastian Hobert}, \bibinfo{person}{Antti Salovaara}, \bibinfo{person}{Matthias S{\"o}llner}, {and} \bibinfo{person}{Jan~Marco Leimeister}.} \bibinfo{year}{2020}\natexlab{}.
\newblock \showarticletitle{Sara, the lecturer: Improving learning in online education with a scaffolding-based conversational agent}. In \bibinfo{booktitle}{\emph{Proceedings of the 2020 CHI conference on human factors in computing systems}}. \bibinfo{pages}{1--14}.
\newblock


\bibitem[Wiratunga et~al\mbox{.}(2024)]%
        {wiratunga2024cbr}
\bibfield{author}{\bibinfo{person}{Nirmalie Wiratunga}, \bibinfo{person}{Ramitha Abeyratne}, \bibinfo{person}{Lasal Jayawardena}, \bibinfo{person}{Kyle Martin}, \bibinfo{person}{Stewart Massie}, \bibinfo{person}{Ikechukwu Nkisi-Orji}, \bibinfo{person}{Ruvan Weerasinghe}, \bibinfo{person}{Anne Liret}, {and} \bibinfo{person}{Bruno Fleisch}.} \bibinfo{year}{2024}\natexlab{}.
\newblock \showarticletitle{CBR-RAG: case-based reasoning for retrieval augmented generation in LLMs for legal question answering}. In \bibinfo{booktitle}{\emph{International Conference on Case-Based Reasoning}}. Springer, \bibinfo{pages}{445--460}.
\newblock


\bibitem[Wu et~al\mbox{.}(2022)]%
        {wu2022ai}
\bibfield{author}{\bibinfo{person}{Tongshuang Wu}, \bibinfo{person}{Michael Terry}, {and} \bibinfo{person}{Carrie~Jun Cai}.} \bibinfo{year}{2022}\natexlab{}.
\newblock \showarticletitle{Ai chains: Transparent and controllable human-ai interaction by chaining large language model prompts}. In \bibinfo{booktitle}{\emph{Proceedings of the 2022 CHI conference on human factors in computing systems}}. \bibinfo{pages}{1--22}.
\newblock


\bibitem[Yan et~al\mbox{.}(2023)]%
        {yan2023xcreation}
\bibfield{author}{\bibinfo{person}{Zihan Yan}, \bibinfo{person}{Chunxu Yang}, \bibinfo{person}{Qihao Liang}, {and} \bibinfo{person}{Xiang'Anthony' Chen}.} \bibinfo{year}{2023}\natexlab{}.
\newblock \showarticletitle{XCreation: A Graph-based Crossmodal Generative Creativity Support Tool}. In \bibinfo{booktitle}{\emph{Proceedings of the 36th Annual ACM Symposium on User Interface Software and Technology}}. \bibinfo{pages}{1--15}.
\newblock


\bibitem[Yang et~al\mbox{.}(2015)]%
        {yang2015uncovering}
\bibfield{author}{\bibinfo{person}{Seungwon Yang}, \bibinfo{person}{Carlotta Domeniconi}, \bibinfo{person}{Matt Revelle}, \bibinfo{person}{Mack Sweeney}, \bibinfo{person}{Ben~U Gelman}, \bibinfo{person}{Chris Beckley}, {and} \bibinfo{person}{Aditya Johri}.} \bibinfo{year}{2015}\natexlab{}.
\newblock \showarticletitle{Uncovering trajectories of informal learning in large online communities of creators}. In \bibinfo{booktitle}{\emph{Proceedings of the Second (2015) ACM Conference on Learning@ Scale}}. \bibinfo{pages}{131--140}.
\newblock


\bibitem[Zhang et~al\mbox{.}(2024)]%
        {zhang2024mathemyths}
\bibfield{author}{\bibinfo{person}{Chao Zhang}, \bibinfo{person}{Xuechen Liu}, \bibinfo{person}{Katherine Ziska}, \bibinfo{person}{Soobin Jeon}, \bibinfo{person}{Chi-Lin Yu}, {and} \bibinfo{person}{Ying Xu}.} \bibinfo{year}{2024}\natexlab{}.
\newblock \showarticletitle{Mathemyths: leveraging large language models to teach mathematical language through Child-AI co-creative storytelling}. In \bibinfo{booktitle}{\emph{Proceedings of the CHI Conference on Human Factors in Computing Systems}}. \bibinfo{pages}{1--23}.
\newblock


\end{thebibliography}
\end{document}